\documentclass[12pt]{article}
\usepackage{amsmath}
\usepackage{graphicx}
\usepackage[round]{natbib}
\usepackage{url} 
\usepackage{xcolor}

\newcommand{\blind}{0}

\usepackage[english]{babel}
\usepackage{amsthm}
\usepackage{amsmath,amssymb,amsfonts,textcomp}
\usepackage{natbib} 
\usepackage{bm}
\usepackage{algorithm}
\usepackage{algpseudocode}
\usepackage{booktabs}
\usepackage{multirow}
\usepackage{graphicx}
\usepackage{subfigure}
\usepackage{dsfont}
\usepackage{url}
\usepackage[]{appendix} 
\usepackage{makecell}
\usepackage{pdflscape}
\usepackage{rotating}

\usepackage{hyperref}

\usepackage[T1]{fontenc}
\usepackage{natbib}

\newcommand{\bt}[1]{\textcolor{black}{#1}} 

\newtheorem{teo}{Theorem}

\usepackage{algorithm}
\usepackage{algpseudocode}
\usepackage{enumitem}
\usepackage{bm}
\usepackage{bbm}

\begin{document}

%
\newcommand{\diff}{\text{d}}
\newcommand{\mC}{\mathcal C}
\newcommand{\mO}{\mathcal O}
\newcommand{\mH}{\mathcal H}
\newcommand{\mG}{\mathcal G}
\newcommand{\mV}{\mathcal V}
\newcommand{\mE}{\mathcal E}
\newcommand{\E}{\mathbb{E}}
\newcommand{\V}{\mathbb{V}}

\newcommand{\ind}{\perp\!\!\!\!\perp} 

\newcommand{\tr}[1]{\mathrm{tr}(#1)}
\newcommand{\diag}[1]{\mathrm{diag}(#1)}

\def\spacingset#1{\renewcommand{\baselinestretch}%
{#1}\small\normalsize} \spacingset{1}


\if0\blind
{
  \title{\bf Gaussian Graphical Models for Partially Observed Multivariate Functional Data}
  \author{Marco Borriero 
  \\
    Department of Statistics, Computer Science and Applications,\\  University of Florence, Italy\\
    \\
    Luigi Augugliaro \\
    Department of Economics, Business and Statistics,\\ University of Palermo, Italy\\
    \\
     Gianluca Sottile \\
    Department of Economics, Business and Statistics,\\ University of Palermo, Italy\\
    \\
    Veronica Vinciotti \\
    Department of Mathematics, University of Trento, Italy}
    \date{}
  \maketitle
} \fi

\if1\blind
{
  \bigskip
  \bigskip
  \bigskip
  \begin{center}
    {\LARGE\bf Title}
\end{center}
  \medskip
} \fi

\bigskip
\begin{abstract}
In many applications, the variables that characterize a stochastic system are measured along a second dimension, such as time. This results in multivariate functional data and the interest is in describing the statistical dependencies among these variables. It is often the case that the functional data are only partially observed. This creates additional challenges to statistical inference, since the functional principal component scores, which capture all the information from these data, cannot be computed. Under an assumption of Gaussianity and of partial separability of the covariance operator, we develop an Expectation-Maximization (EM)-type algorithm for penalized inference of a functional graphical model from multivariate functional data which are only partially observed. A simulation study and an illustration on \bt{environmental, social and governance (ESG) data} show the potential of the proposed method.
\end{abstract}

\noindent%
\vfill

\newpage
\spacingset{1.75} 
\section{Introduction}
In recent years, the study of data in the form of functions or curves, referred to as functional data analysis \citep{Ramsay,Ferraty,horvath2012inference,Hsing}, has received much attention, with applications in a variety of fields, from medicine~\citep{LiEtAl_JASA_18,ZhaoEtAl_EJS_24} to economics~\citep{liebl2013,GOLDBERG2014}.  By their own nature, functional data are high dimensional. Therefore, dimensionality reduction techniques are often needed in order to make them more tractable. The most common is functional principal components analysis (FPCA),
which uses the Karhunen-Loève theorem to represent functional data as an infinite linear combination of a given basis system, with coefficients called functional principal components scores, or simply scores. The advantage of this representation is that the scores are random variables which contain all the information about the data. Moreover, FPCA gives automatically an order of importance to the sequence of scores, with the first one containing the majority of the information and then proceeding in a decreasing fashion.

In this work, we focus on multivariate functional data in which every observation is represented by different curves. In particular, we consider the problem of modelling the dependences between the curves of the multivariate process through a functional Gaussian graphical model (FGGM)\citep{ZhuEtAl_JMLR_16, Lee2023, QiaoEtAl_JASA_19, ZapataEtAl_BioK_22}. For this purpose, a crucial role is played by the covariance operator, which is an extension of the covariance matrix of a multivariate random variable to the functional case. However, the non invertibility of this operator makes the inference of a \bt{FGGM} more challenging than in the non-functional context.

Two fundamental contributions in this direction are provided by \cite{QiaoEtAl_JASA_19} and \cite{ZapataEtAl_BioK_22}, who address the problem in two different ways. \cite{QiaoEtAl_JASA_19} approximate the functional data with a truncated Karhunen-Loève expansion, obtaining in this way a finite dimensional precision matrix of the scores that is used to recover the edge set of the functional graphical model. 
\bt{The approach of \cite{QiaoEtAl_JASA_19}, which has been recently extended also to the case of dynamic networks~\citep{QiaoEtAl_BioK_20},  can recover the true edge set only under the restrictive assumption that each random function has a finite expansion. To overcome this limitation, \citet{ZapataEtAl_BioK_22} propose an additive hypothesis on the covariance operator, called partial separability
}. This gives rise to a novel version of the Karhunen-Loève expansion for multivariate functional data. Thanks to this hypothesis, the edge set of a \bt{FGGM} can be obtained from the precision matrices of the scores.
More recently, \cite{fici2025functional} have extended this second method to include a dependence of the functional curves on external covariates, while \cite{ZhaoEtAl_EJS_24} have proposed a functional generalization of the neighborhood selection method~\citep{MeinshausenEtAl_AoS_06} which avoids the definition of the precision operator and the partial separability assumption on the covariance operator but focuses primarily on graph selection.

To date, the literature on \bt{FGGM} has considered only the setting in which functional data are completely observed.  However, functional data are often recorded with missing values. This is not the case of sparse functional data, i.e., when few observations are available in the considered time domain \citep{Yao,James}.
It is rather the case when, for instance, the technological device that is used for taking measurements stops working properly for a certain period of time or when the measures are recorded but considered as not plausible for a variety of reasons. Situations like this give rise to functional data that are partially observed. In these cases, one may be interested in  reconstructing the trajectories of the curves in the missing domains and in studying the dependence structure among the curves.

There are approaches in the literature for partially observed functional data in the univariate case. In particular, \cite{Kraus} propose a methodology for the functional completion of univariate functional data through a functional linear regression with the observed data as predictors, while \cite{Delaigle_and_Hall} approximate fragmented functional data with a Markov chain model.
Finally, \cite{Liebl2} define a new reconstruction operator, which belongs to a broader class than the class of linear operators and that is optimal in the sense that the reconstructed curve in the missing domain preserves the continuity at the extremes of the observed intervals. 


\bt{In this paper, we contribute to both the literature on \bt{FGGMs} and that on partially observed functional data by proposing a novel approach that, under the assumption of partial separability of the covariance operator, simultaneously reconstructs the missing fragments of the random functions and estimates their conditional dependence structure. At the core of our proposal lies a novel EM-type algorithm, with an E-step that implements a multivariate generalization of the procedure of \cite{Kraus} to reconstruct the missing fragments of the functions, and an M-step that uses the approach of~\cite{ZapataEtAl_BioK_22} to consistently estimate the edge set associated with the Gaussian graphical model.}

The paper is organized as follows. \bt{In Section~\ref{subsec:esg_motivating_example}, we present a motivating example, while in} Section~\ref{sec:fggm} we introduce multivariate functional data and \bt{FGGMs}. In Section~\ref{sec:inference}, we discuss the methodology for the inference of a \bt{FGGM} from partially observed multivariate data, and present the technical details of the multivariate score imputation and the proposed EM-type algorithm. Section~\ref{sec:simulation} presents a simulation study, showing the performance of the method \bt{across a number of settings} and compared to existing methods. Finally, Section ~\ref{sec:esg_application} concludes with an illustration of the methodology on \bt{Environmental, Social and Governance (ESG) data from the World Bank \citep{WorldBankFramework2026}.}

\bt{\section{Motivating Example}
\label{subsec:esg_motivating_example}
Sovereign ESG data measure the Environmental, Social, and Governance (ESG) performance of a country. The data are made available by the World Bank through the Sovereign ESG Data Portal \citep{WorldBankFramework2026} and  help investors and policymakers evaluate a nation's long-term sustainability risks, development outcomes, and capacity to manage its sovereign debt. The data are a natural example of multivariate functional data since measurements for each country are made over time across a number of different dimensions of performance of a country, such as in relation to environmental quality, social inclusion, institutional capacity and structural transformation.} 

\bt{In the empirical literature, sovereign ESG data are typically analyzed  through highly aggregated scores built from ad-hoc weighting and normalization schemes. These  scores have been shown to introduce biases,  whereby richer countries receive systematically better ESG assessments independently of the specific sustainability mechanism under study \citep{IMF2024SovereignESG}. Moreover, a strong dependence structure between the different  sovereign ESG indicators is expected. However, this is typically ignored both in the construction of aggregate scores and in the imputation of the large percentage of missing that characterize these data. The framework proposed in this paper is therefore particularly suited to this setting. We model  sovereign
ESG data as a multivariate system of partially observed trajectories of country performances over time. The procedure will allow both the reconstruction of the  dependence structure across the different dimensions, which is more transparent than
an aggregate score and better suited to policy interpretation, and a more accurate imputation of missing data that accounts for this dependence structure.}

\begin{figure}[!htbp]
	\centering
    \begin{minipage}{0.45\textwidth}
        \centering
        \includegraphics[width=\linewidth]{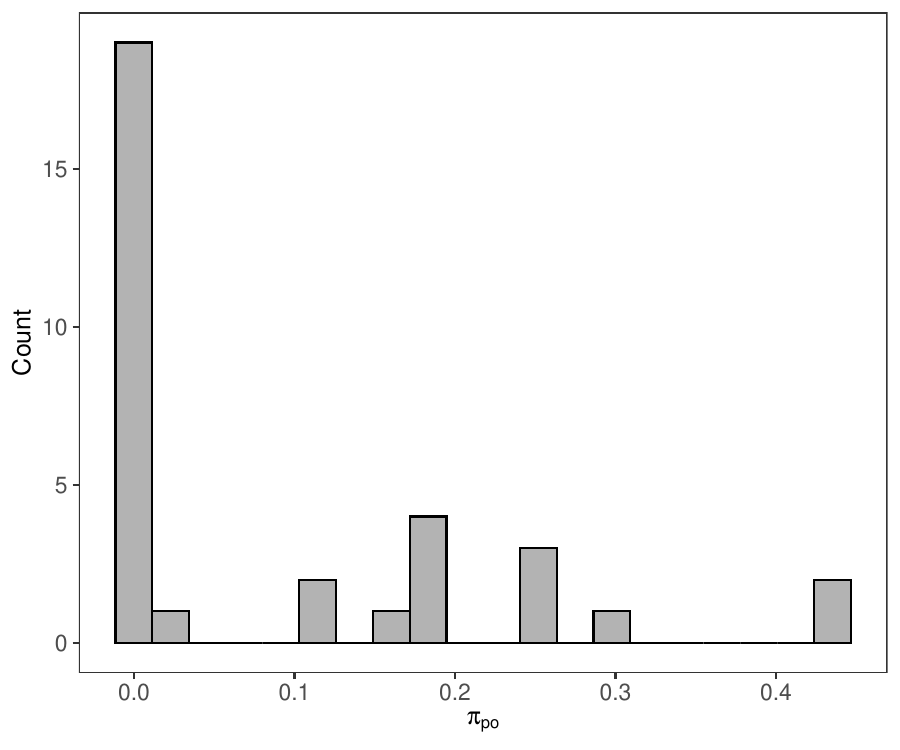}
    \centering (a)
    \end{minipage}
    \hfill
    \begin{minipage}{0.51\textwidth}
        \centering
        \includegraphics[height=6cm]
        {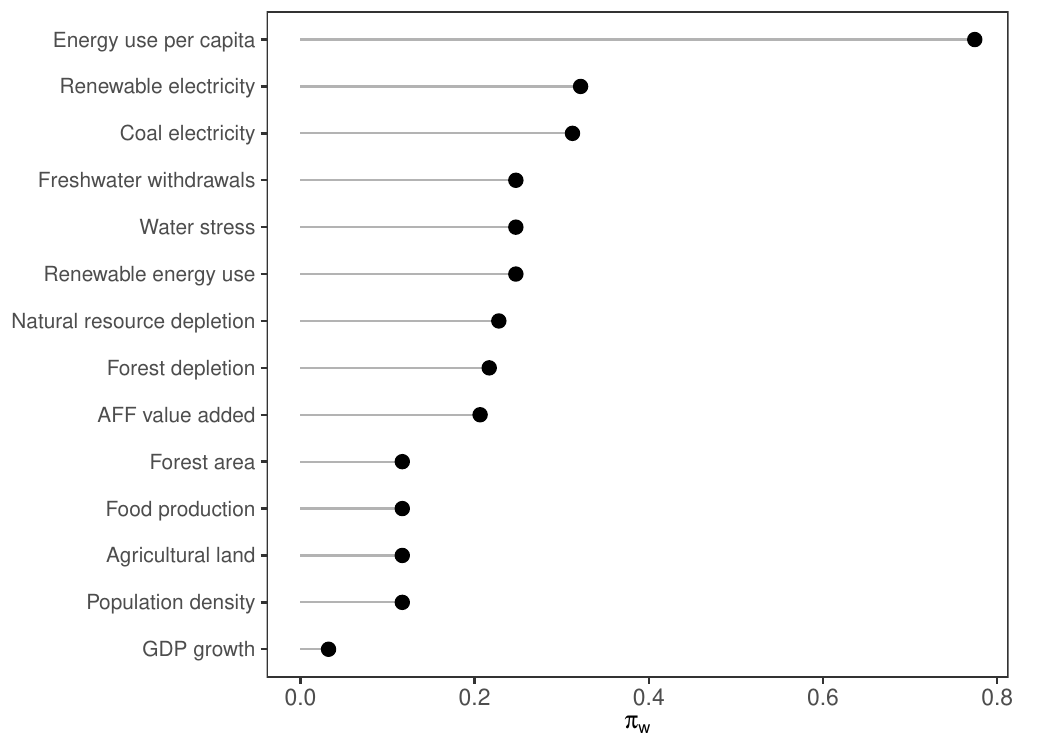}
    \centering (b)
    \end{minipage}
	\caption{\small Summary of missingness patterns across the ESG indicators used for the empirical illustration: (a)
     number of indicators that have a certain proportion of partially observed curves across all countries ($\pi_{\mathrm{po}}$); (b) 
average proportion of unobserved time points ($\pi_w$) for the indicators that are partially observed.}
	\label{fig:summary_missing}
\end{figure}

\bt{The data portal provides information for $214$ countries and $195$ indicators for the years between $1960$ and $2025$. For the analysis, we consider the $p=33$ indicators reported in Table~\ref{tab:esg_indicators}, which belong to the $78$ indicators of the core sovereign ESG framework \citep{WorldBankFramework2026}. The indicators are evaluated on the $31$ years between $1990$ and $2020$ and for $n=131$ countries. This is a combination that ensured sufficient coverage across indicators, years and countries. Table~\ref{tab:esg_indicators} groups these indicators according to the main empirical blocks defined by the World Bank sovereign ESG framework, showing how the chosen indicators jointly cover environmental, social and governance performance measures of a country. Figure~\ref{fig:summary_missing} summarizes the level of missingness in the data. In particular, Figure~\ref{fig:summary_missing}a reports the number of indicators that have a certain proportion of partially observed curves across all countries, which we later refer to as $\pi_{\mathrm{po}}$. The figure shows how there is a number of indicators for which there is complete data, but others that have at least one missing value on up to 50\% of countries. Figure~\ref{fig:summary_missing}b reports the average proportion of unobserved time points, later referred to as $\pi_w$, for the indicators that are partially observed. This shows how, for some indicators, there can be up to $80\%$ of missing data along the temporal trajectory. 
}
\begin{table}[!htb]
	\scriptsize
	\centering
	\caption{\small World Bank sovereign ESG indicators used for the empirical illustration. The ESG pillar and group classifications follow the World Bank sovereign ESG framework \citep{WorldBankFramework2026}.}
	\label{tab:esg_indicators}
    \scriptsize
    \setlength{\tabcolsep}{3pt}      
    \renewcommand{\arraystretch}{0.92} 
	\resizebox{\textwidth}{!}{
    \begin{tabular}{p{2.2cm} p{4.5cm} p{4.0cm} p{5.85cm}}
		\toprule
		\textbf{ESG pillar} & \textbf{Group} & \textbf{Indicator} & \textbf{Brief description} \\
		\midrule
		
		Environment & Emissions \& pollution & PM2.5 exposure & Ambient air pollution exposure \\
		 & & CO$_2$ total & Aggregate carbon dioxide emissions \\
		 &  & CO$_2$ per capita & Per-capita carbon dioxide emissions \\
		 &  & CH$_4$ total & Aggregate methane emissions \\
		 &  & N$_2$O total & Aggregate nitrous oxide emissions \\
		 &  & GHG total & Aggregate greenhouse gas emissions \\
		 &  & GHG per capita & Per-capita greenhouse gas emissions \\
		\cmidrule{2-4}
		 & Energy use \& security & Coal electricity & Electricity generated from coal \\
		 &  & Energy use per capita & Per-capita energy consumption \\
		 &  & Renewable electricity & Share of electricity from renewables \\
		 &  & Renewable energy use  & Share of final energy from renewables \\
		\cmidrule{2-4}		
		 & Climate risk \& resilience & Cooling degree days & Exposure to cooling demand \\
		 &  & Heat stress & Exposure to extreme heat conditions \\
		 &  & Heating degree days & Exposure to heating demand \\
		 &  & Water stress & Pressure on freshwater availability \\
		 &  & Population density & Population pressure on land \\
		 &  & SPEI & Hydro-climatic dryness/wetness index \\
		\cmidrule{2-4}
		 & Food Security & Agricultural land & Share of land used for agriculture \\
		 &  & Agriculture, forestry and fishing (AFF) value added & Economic weight of agriculture, forestry and fishing \\
		 &  & Food production & Food production dynamics \\
		\cmidrule{2-4}
		 & Natural capital use \& management & Natural resource depletion & Depletion of natural resource rents \\
		 &  & Forest depletion & Net depletion of forest resources \\
		 &  & Freshwater withdrawals & Total freshwater withdrawals \\
		 &  & Forest area & Share of land covered by forest \\
		\midrule
		Social & Demography & Fertility & Total fertility rate \\
		 &  & Life expectancy & Life expectancy at birth \\
		 &  & Population aged 65+ & Population ageing \\
		\cmidrule{2-4}
		 & Employment & Labor force participation & Total labor market participation \\
		\cmidrule{2-4}
		 & Health \& Nutrition & Under-5 mortality & Child mortality outcome \\
		 &  & Overweight prevalence & Adult overweight prevalence \\
		\midrule		
		Governance & Economic Environment & GDP growth & Annual economic growth \\
		\cmidrule{2-4}
		 & Gender & Female-to-male labor ratio & Gender balance in labor participation \\
		\cmidrule{2-4}
		 & Stability \& Rule of Law & Net migration & Net migration balance \\		
		\bottomrule
	\end{tabular}}
\end{table}

\section{Functional Gaussian Graphical Models} \label{sec:fggm}


\subsection{Notation}
We denote by $L^2_{[0,1]}$  the separable Hilbert space of square-integrable real-functions defined on $[0,1]$, which is equipped with the standard inner product $\langle f, g \rangle = \int_0^1 f(t)g(t)\diff t$, for any $f,g \in L^2_{[0,1]}$, and induced $L_2$-norm $\|f\| = \langle f, f \rangle^{1/2}$. For any given subset of the closed interval $[0, 1]$, say $S$, the restrictions of the inner product and the norm are denoted by $\langle f^S, g^S\rangle = \int_S f(t)g(t)\diff t$ and $\|f^S\| = \langle f^S, f^S\rangle^{1/2}$, respectively. The $p$-fold Cartesian product of the spaces $L^2_{[0,1]}$ is denoted by $\left(L^{2}_{[0,1]}\right)^p$ and the inner product and norm are defined as $\langle \bm{f},\bm{g} \rangle= \sum_{j=1}^{p} \langle f_{j},g_{j} \rangle$ and $\|\bm f\| = \langle \bm{f},\bm{f} \rangle^{1/2}$, respectively. For any given set of $p$ subsets of $[0, 1]$, say $\bm S = \{S_j\}_{j = 1}^p$, we denote by $\langle \bm{f^S}, \bm{g^S} \rangle = \sum_{j = 1}^p\langle f^{S_j}_j, g^{S_j}_j\rangle$ and $\|\bm{f^S} \|$  the restriction of the inner product and norm, respectively. Finally, given a $p\times p$-dimensional matrix of functions, say $\bm M = \{M_{hk}\}$, with $M_{hk}\in L^2_{[0, 1]}$, we denote by $\bm M_k = \{M_{hk}\}\in\left(L^2_{[0, 1]}\right)^p$  the $k$th column of $\bm M$ and, by $\langle \bm M, \bm f\rangle\in \left(L^2_{[0, 1]}\right)^p$ the $p$-dimensional vector whose \bt{$k$th component} is equal to $\langle \bm M_k, \bm f\rangle$, for any given $\bm f \in\left(L^2_{[0, 1]}\right)^p $. Given two matrices of functions, with dimension $p\times p$, say $\bm M$ and $\bm N$, we denote by $\langle \bm M, \bm N\rangle$  the $p\times p$ dimensional matrix with generic entry $\langle \bm M_h, \bm N_k\rangle$.


\subsection{\bt{Review of related works}}
Multivariate functional data are a random sample from a multivariate process $\{\bm X(t)\in\mathds R^p\;:\;t\in[0, 1]\}$, which is assumed to be zero-mean such that $\bm X\in \left(L^{2}_{[0,1]}\right)^p$ almost surely and $\E(\|\bm X \|_p)<\infty$. Functional Gaussian graphical models~\citep{ZhuEtAl_JMLR_16} represent a new class of graphical models which extends  the classical Gaussian graphical models~\citep{Lauritzen_book_96} to infinite dimensional spaces. In particular,  $\bm X$ is assumed to follow a multivariate Gaussian process and the conditional independence structure among the $p$ random functions is encoded by an undirected graph $\mG = \{\mV, \mE\}$, i.e., the pair $(h, k)$ belongs to the edge set $\mE$ if and only if $X_{h}\ind X_{k}\mid\bm X_{\mV\setminus\{h, k\}}$. 

As in the finite dimensional case, a multivariate Gaussian process is uniquely determined by the covariance operator $\mC: \left(L^{2}_{[0,1]}\right)^p\rightarrow \left(L^{2}_{[0,1]}\right)^p$, with $C_{hk}(s, t) = \E(X_h(s)\, X_k(t))$ the cross-covariance function between $X_h$ and $X_k$ \citep{Prato_Book_06}. Following~\cite{ChiouEtAl_StatSinica_14}, the operator $\mC$ is formally defined as the integral operator with covariance kernel $\bm C(s, t) = \{C_{hk}(s, t)\}$, i.e.,
\begin{equation}\label{eqn:CovOp}
\mC\bm f(s) = \langle \bm C(s, \cdot), \bm f \rangle =  %
\begin{pmatrix}
\langle\bm C_1(s, \cdot), \bm f\rangle\\
\vdots \\
\langle\bm C_p(s, \cdot), \bm f\rangle
\end{pmatrix},\qquad\forall \bm f\in\left(L^{2}_{[0,1]}\right)^p,
\end{equation}
where $\langle\bm C_k(s, \cdot), \bm f\rangle = \sum_{h = 1}^p\langle C_{hk}(s, \cdot),  f_h\rangle$. Using the operator~(\ref{eqn:CovOp}), we denote by $MGP(\mC)$  a multivariate Gaussian process and by $\{MGP(\mC), \mG\}$ the corresponding \bt{FGGM}. 

As discussed in~\cite{QiaoEtAl_JASA_19}, although the functional setting is more complicated, some of the main properties of the standard Gaussian graphical model are inherited in an infinite dimensional setting. Specifically, letting $\bm{X}_{\setminus(h,k)} = \{X_m\,:\,m\in\mV\setminus(h,k)\}$, it is possible to show that the edge set can be defined in terms of the conditional cross-covariance function $K_{hk}(s,t) = Cov(X_h(s), X_k(t) \mid \bm{X}_{\setminus(h,k)})$, which represents the covariance between $X_h(s)$ and $X_k(t)$ conditional on the remaining $p-2$ random functions. In particular, it follows that:
\[
\mE= \{(h,k) : K_{hk}(s,t) \neq 0 \text{ for some } s \text{ and } t \in [0,1], \; (h,k)\in\mV^2, \; h \neq k \}.
\] 

To estimate the edge set from multivariate functional data, \cite{QiaoEtAl_JASA_19} propose a two-steps procedure. First, they use a finite approximation of each random function. This is formally obtained using the first $L_h$ terms of the univariate Karhunen-Loève expansion $X_{h}(t)= \sum_{l=1}^{+\infty}  \zeta_{hl} \varphi_{hl}(t)$, where $\{\varphi_{hl} \}_{l=1}^{+ \infty}$ are the eigenfunctions and form a complete orthonormal basis system (CONS) for $L^2_{[0, 1]}$. Under the Gaussianity assumption of the process, one can show that $\zeta_{hl}\sim N(0, \lambda_{hl})$, with $\lambda_{h1}\ge\lambda_{h2}\ge\dots\ge0$. \bt{Moreover, $\zeta_{hl}$ is independent of $\zeta_{h'l'}$ for 
$l\ne l'$.} Since the random vector $\bm\zeta = (\zeta_{11},\ldots,\zeta_{1L_1},\ldots,\zeta_{p1},\ldots,\zeta_{pL_p})^\top$ follows a multivariate Gaussian distribution whose precision matrix has a specific block structure, the authors propose a generalization of the graphical lasso (glasso) estimator~\citep{YuanEtAl_BioK_07}, called functional graphical lasso, in order to encourage blockwise sparsity in the resulting precision matrix. As shown in \cite{QiaoEtAl_JASA_19}, the approach can recover the  edge set $\mE$ only under the restrictive assumption that each random function has a finite representation.

The reason for this theoretical restriction lies in the observation that the covariance operator is compact and therefore not invertible~\citep{Hsing}.  Consequently, the connection between conditional independence and an inverse covariance operator is lost, as the latter does not exist. To overcome this methodological problem, ~\cite{ZapataEtAl_BioK_22} introduce a specific assumption on the structure of the covariance operator called partial separability. Formally, the covariance operator $\mC$ is said to be partially separable if there exists a CONS of $L^2_{[0, 1]}$, denoted by $\{\varphi_{l} \}_{l=1}^{+ \infty}$, and if, for each $l\in\mathds N$, there exists an orthonormal matrix $\bm Q_l = (\bm q_{1l}\mid\dots\mid\bm q_{pl})$ of dimension $p\times p$ such that: 
\[
\mC(\bm q_{hl}\varphi_l)(s) = \lambda_{hl} \bm q_{hl}\varphi_l(s).
\]

The main advantage of this assumption is that it implies the following multivariate Karhunen-Loève expansion: 
\begin{equation}\label{eqn:MKL}
\bm X(t) = \sum_{l = 1}^{\infty} \bm \xi_l\varphi_l(t),
\end{equation}
where, under the assumption of a Gaussian process, $\bm\xi_l$ are \bt{mutually independent} Gaussian random vectors with $\E(\bm\xi_l) = \bm 0$, $\E(\bm\xi_l\bm\xi_l^\top) = \Sigma_l = \bm Q_l^\top\Lambda_l\bm Q_l$ and $\Lambda_l = \diag{\lambda_{1l},\ldots,\lambda_{pl}}$. The expansion in~(\ref{eqn:MKL}) is assumed to be ordered according to the decreasing value of $\tr{\Sigma_l}$. Although the assumption of partial separability relates the basis $\{\varphi_l\}_{l = 1}^{+\infty}$ to the covariance operator $\mC$, it is possible to show that its theoretical properties, such as uniqueness and optimality, depend on the covariance operator $\mH = \frac{1}{p}\sum_{h = 1}^p\mC_{hh}$, which admits, under the assumption of partial separability, the following spectral decomposition:
\[
\mH = \sum_{l = 1}^{+\infty} \lambda_l\varphi_l\otimes\varphi_l, 
\]
where $\lambda_l = p^{-1}\tr{\Sigma_l}$ and the tensor product is the operator $(\varphi_l\otimes\varphi_l)(\cdot) = \langle\varphi_l,\cdot\rangle\varphi_l$ on $L^2_{[0, 1]}$.

The advantage of the representation~(\ref{eqn:MKL}) is related to the form of the edge set of the \bt{FGGM}. Indeed, letting $\Theta_l = \Sigma_l^{-1}$ and $\theta_{hkl}$ the $(h,k)$ entry of $\Theta_l$, if the covariance operator $\mathcal{C}$ is partially separable, then there exists a sequence of edge sets $\{\mE_l\}_{l=1}^{+ \infty}$, defined as $\mE_l = \{ (h,k) : \theta_{hkl} \neq 0, \, h \neq k \}$, such that the edge set of the \bt{FGGM} is given by $\mE = \bigcup_{l=1}^{+ \infty}\mE_l$. Notice that $\mE_l$ can be seen as the edge set of the Gaussian graphical model associated with $\bm{\xi}_{l} \sim N(\textbf{0}, \Sigma_l)$. This shows that, under the hypothesis of partial separability, the edge set of the \bt{FGGM} can be recovered from the sequence of precision matrices $\{ \Theta_l\}_{l=1}^{+ \infty}$. Moreover, \citet{ZapataEtAl_BioK_22} prove that the edge selection procedure is consistent. That is, for a fixed $n$, there exists a $L=L_p \in \mathbb{N}$ such that $\mE = \bigcup_{l=1}^{L_p} \mE_l$ and $L_p$ diverges with $n$ only if $p$ diverges. Since the precision matrices $\Theta_l = \Sigma_l^{-1}$ contain all the necessary information to estimate the relevant terms of a \bt{FGGM},  \cite{ZapataEtAl_BioK_22} propose to use an extension of the glasso estimator, called joint graphical lasso estimator~\citep{Danaher}, which can recover, under specific assumptions, the true graph $\mG$. 


\section{Inference from partially observed functional data}\label{sec:inference}

In this section, we first formalize the inferential problem under \bt{fragmented functional data}, by defining a suitable objective function. We then embed this in an Expectation-Maximization (EM)-type algorithm. A crucial step of the algorithm is the imputation of the scores, which we discuss in the next subsection, before concluding with the computational aspects of the algorithm.

\subsection{Expected log-likelihood function}\label{sec:EStep}
Let $\bm X\sim MGP(\mC)$ be a $p$-dimensional vector of random functions following a multivariate Gaussian process with a partially separable covariance operator. For each function, we consider a partition in the domain $[0, 1]$. As we consider the case of partially observed data, the domain of the $j$th random function $X_j$ is split into $O_j$ and $M_j$, representing the portions in which the function is observed and unobserved, respectively. In both cases, $O_j$ and $M_j$ might be given by the union of a finite number of sub-intervals. As in \cite{Kraus}, we further assume that the functional missing data is missing completely at random, which means that observed and missing domains can be thought as fixed subsets. In the remaining part of this paper, we denote by $X_j^{O_j}$ and $X_j^{M_j}$  the restriction of the random function $X_j$ over the subdomains $O_j$ and $M_j$, respectively.

When the functional data are not observed in their entire domain, the entries of the $p$ dimensional random score $\bm{\xi}_l$ of the expansion~(\ref{eqn:MKL}) cannot be computed directly. Formally, each $\xi_{jl}$ can be written as a sum of an observed and a missing term:
\begin{equation}\label{eqn:xiO_xiM}
\xi_{jl} = \langle X_j, \varphi_l \rangle = \langle X_j^{O_j}, \varphi^{O_j}_l \rangle + \langle X_j^{M_j}, \varphi^{M_j}_l \rangle =
\xi_{jl}^o + \xi_{jl}^m, \quad \forall j=1,\dots,p \, , \; \forall l\in\mathds N,
\end{equation}
where $\varphi^{O_j}_l$ and $\varphi^{M_j}_l$ denote the restriction of $\varphi_l$ on $O_j$ and $M_j$, respectively. Since $X_{j}$ is not observed in $M_{j}$,  the term $\xi_{jl}^m$ is missing and, consequently, some imputation is needed in order to complete the random function $X_j$. 

To define a proper objective function by which to estimate the conditional independence structure among the $p$ random functions, we propose to use the first $L$ leading terms of the expansion~(\ref{eqn:MKL}) along with the identity~(\ref{eqn:xiO_xiM}), so that the log-likelihood function can be written as
\begin{equation}\label{eqn:loglik}
\ell(\{\Theta\}) = \sum_{l = 1}^{L}\left\{\frac{1}{2}\log\det \Theta_l - \frac{1}{2}(\bm{\xi^o}_l + \bm{\xi^m}_l)^\top\Theta_l(\bm{\xi^o}_l + \bm{\xi^m}_l)\right\},
\end{equation}
where $\{\Theta\} = \{\Theta_1,\ldots, \Theta_L\}$ denotes the set of $L$ precision matrices. 

We use function~(\ref{eqn:loglik}) to develop an algorithm that follows the rationale of the Expectation-Maximization~(EM) algorithm~\citep{Dempster}. In particular, instead of maximizing function~(\ref{eqn:loglik}), which cannot be computed as it depends on the missing scores, we maximize the conditional expected value of the log-likelihood function given the observed functions $\bm{X^O} = \{X_j^{O_j}\}_{j = 1}^p$:
\[
\E[\ell(\{\Theta\})\mid \bm{X^O}] = \sum_{l = 1}^{L}\left\{\frac{1}{2}\log\det \Theta_l - \frac{1}{2}\E[(\bm{\xi^o}_l + \bm{\xi^m}_l)^\top\Theta_l(\bm{\xi^o}_l + \bm{\xi^m}_l)\mid \bm{X^O}]\right\},
\]
which can be simplified by noting that  \bt{the argument of the exponent can be expanded as follows:}
\begin{multline*}
\E[(\bm{\xi^o}_l + \bm{\xi^m}_l)^\top\Theta_l(\bm{\xi^o}_l + \bm{\xi^m}_l)\mid \bm{X^O}] = \\ \mathrm{tr}\{(\bm{\xi^o}_l)(\bm{\xi^o}_l)^\top\Theta_l\} + \mathrm{tr}\{\bt{\E[(\bm{\xi^m}_l)(\bm{\xi^m}_l)^\top\mid \bm{X^O}]}\,\Theta_l\} + 2 \mathrm{tr}\{\bt{\bm{\mu^{m\mid o}}_l}(\bm{\xi^o}_l)^\top\Theta_l\}.
\end{multline*}
\bt{where $\bm{\mu^{m\mid o}}_l = (\mu^{m\mid o}_{1l},\ldots, \mu^{m\mid o}_{pl})^\top$ denotes the conditional expected value of the missing score $\bm{\xi^m}_l$ given the observed fragment $\bm{X^O}$. To simplify the previous expression we use the approach proposed by~\cite{Guo}, i.e., we use the mean field theory approach~\citep{PetersonEtAl_CompSyst_87} to approximate the conditional second moment with  $\E[\xi^m_{hl}\xi^m_{kl}\mid \bm{X^O}]\approx \mu^{m\mid o}_{hl}\mu^{m\mid o}_{kl}$. As discussed in~\cite{Guo}, this approximation performs well when $\xi^m_{hl}$ and $\xi^m_{kl}$ are close to be conditionally independent, which holds when $\Theta_l$ is sparse and its off-diagonal entries are not too large. In other terms, $\Theta_l$ must be close to the diagonal matrix $D_l =\diag{\Theta_l}$. This can be formalized by the condition $\|D_l^{-1/2}\Omega_l D_l^{-1/2}\|_2 \ll 1$, where $\Omega_l$ is the matrix containing the off-diagonal entries of $\Theta_l$ and $\|\cdot\|_2$ denotes the spectral norm. The approximation proposed by~\cite{Guo} has been used successfully in a number of methodological papers on the inference of Gaussian graphical models from partially observed data (see for example~\cite{AugugliaroEtAl_BioStat_18, AugugliaroEtAl_CompStat_20, AugugliaroEtAl_JSS_23, SottileEtAl_JRSSC_24} and references therein). Finally, substituting $\E[\xi^m_{hl}\xi^m_{kl}\mid \bm{X^O}]$ with $\mu^{m\mid o}_{hl}\mu^{m\mid o}_{kl}$, the conditional expected value of the log-likelihood function can be approximated as follows:}
\[
\E[\ell(\{\Theta\})\mid \bm{X^O}] \approx \sum_{l = 1}^{L}\left\{\frac{1}{2}\log\det \Theta_l - \frac{1}{2}(\bm{\xi^o}_l + \bm{\mu^{m\mid o}}_l)^\top\Theta_l(\bm{\xi^o}_l + \bm{\mu^{m\mid o}}_l)\right\}.
\]
%
%
%
%

Since a maximization of this function will lead to estimated precision matrices with all non-zero entries,  we explore \bt{FGGMs} with different degrees of complexity, i.e., edge sets with varying sparsity levels, by maximizing, similarly to~\cite{ZapataEtAl_BioK_22}, the following penalized expected likelihood
\begin{equation}\label{eqn:Q}
\sum_{l = 1}^{L}\left\{\frac{1}{2}\log\det \Theta_l - \frac{1}{2}(\bm{\xi^o}_l + \bm{\mu^{m\mid o}}_l)^\top\Theta_l(\bm{\xi^o}_l + \bm{\mu^{m\mid o}}_l)\right\} - P_{\bm\gamma}(\{\Theta\}),
\end{equation}
where $P_{\bm\gamma}(\{\Theta\})$ is the group graphical lasso penalty function proposed in~\cite{Danaher}
\begin{equation}\label{eqn:GGL}
P_{\bm\gamma}(\{\Theta\}) = \gamma_1\left\{\gamma_2\sum_{l = 1}^L\sum_{h\ne k}^p|\theta_{hkl}| + (1 - \gamma_2) \sum_{h\ne k}^p \left(\sum_{l = 1}^L \theta_{hkl}^2\right)^{1/2}\right\}.
\end{equation}
In particular, $\gamma_1$ is a non-negative tuning parameter that controls the overall penalty level, while $\gamma_2\in[0, 1]$ distributes the penalty between the two penalty functions. 

Table~\ref{tbl:PseudoCode} summarizes the main steps of the proposed EM-type algorithm for the optimization of \eqref{eqn:Q}. \bt{As one can see, the proposed algorithm involves estimating the eigenfunctions only at the initial stage, i.e., Step 1, and subsequently treats them as fixed throughout the entire estimation process. The reason lies in the fact that updating $\varphi_l$ within the optimization process using the eigenfunction equation, i.e., using the equation $\langle\mH(s,\cdot),\varphi_l\rangle = \lambda_l\varphi_l(s)$,  would not result in a change to the initial estimates. 
This follows from the assumption of partial separability, which implies that the $p$ random functions share the same system of basis functions and therefore that an initial estimate from the entire functional data is optimal. } Finally, the pseudo-code reveals also that a crucial step is played by the imputation of the missing scores. This aspect will be discussed in the next section.

\begin{table}[!ht]
\caption{Pseudo-code of the proposed EM-type algorithm\label{tbl:PseudoCode}}
\begin{center}
\begin{tabular}{cl}
\hline
Step & Description\\
1 & compute $\varphi_l$ as solution of the following identity\\
& $\langle\mH(s,\cdot),\varphi_l\rangle = \lambda_l\varphi_l(s)$\\
2 & choose the number of $L$ terms to approximate expansion~(\ref{eqn:MKL})\\
3 & repeat\\
4 & \qquad for $l = 1$ to $L$ \\
5 & \qquad\qquad let $\bm{\widehat{\mu}^{m\mid o}}_l$ be an estimate of $\bm{\mu^{m\mid o}}_l$\\
6 & \qquad\qquad let $\bm{\hat\xi}_l = \bm{\xi^o}_l + \bm{\widehat{\mu}^{m\mid o}}_l$\\
7 & \qquad and for\\
8 & \qquad estimate the precision matrices \\
  & \qquad $\{\widehat{\Theta}_{\bm\gamma}\} = \arg\max \sum_{l = 1}^{L}\left\{\frac{1}{2}\log\det \Theta_l - \frac{1}{2}\bm{\hat\xi}_l^\top\Theta_l\bm{\hat\xi}_l\right\} - P_{\bm\gamma}(\{\Theta\})$ \\
9 & until a convergence criterion is met\\
\hline
\end{tabular}
\end{center}
\end{table}

\subsection{Multivariate imputation of the missing  scores}
For the imputation of the scores, we generalize the approach of \cite{Kraus} to the multivariate case. In this way, the imputation of missing data exploits the dependence structure among the $p$ random functions.
In terms of the mean-squared prediction error, the $j$th entries of the missing score $\bm{\xi^m}_l$, denoted by $\xi_{jl}^{m}$, can be imputed using $\mathbb{E}[{\xi}_{jl}^{m} \mid \bm{X^O}] = \mu_{jl}^{m\mid o}$. Since $\mu_{jl}^{m\mid o}$ could be a non-linear functional of $\bm{X^O}$, we propose to model $\mu_{jl}^{m\mid o}$ using a continuous linear functional of the observed functions. Then, by the Riesz representation theorem, we have $\mu_{jl}^{m\mid o} = \sum_{i = 1}^p \langle \beta_{ijl}, X^{O_i}_i\rangle = \langle \bm \beta_{jl}, \bm{X^O}\rangle$, where $\bm \beta_{jl} = \{\beta_{ijl}\}_{i = 1}^p$ is a $p$-dimensional vector of functions with $\beta_{ijl}\in L^2_{O_i}$. In matrix form, denoting by $\bm B_l = (\bm\beta_{1l}\mid\cdots\mid\bm\beta_{pl})$ the matrix of regression coefficient functions, the conditional expected value of $\bm{\xi^m}_l$ is modelled by 
\[
\bm{\mu^{m\mid o}}_l = \langle \bm B_l, \bm{X^O}\rangle = 
\begin{pmatrix}
\langle \bm{\beta}_{1l}, \bm{X^O} \rangle \\
\vdots\\
\langle \bm{\beta}_{pl}, \bm{X^O} \rangle
\end{pmatrix}. 
\]

Next, we discuss how to define the infinite-dimensional optimization problem by which to estimate $\bm B_l$. Under the assumption that $\bm{\xi^m}_l  \sim N_p(\langle\bm B_l, \bm{X^O}\rangle, {\Sigma}_l^{m})$, where ${\Sigma}_l^{m}$ denotes the covariance matrix of the vector $\bm{\xi^m}_l$, the optimal linear functional can be defined as the solution of the following minimization problem
\begin{equation}\label{eqn:Prob_Bhat}
\min_{\bm B_l} \frac{1}{2} \mathbb{E} \left\{ (\bm{\xi^m}_l - \langle \bm B_l, \bm{X^O}\rangle)^\top {\Theta}_l^{m} (\bm{\xi^m}_l - \langle \bm B_l, \bm{X^O}\rangle)\right\},
\end{equation}
where ${\Theta}_l^{m}$ is the inverse of ${\Sigma}_l^{m}$. 
The following theorem discusses the solution to this optimization problem.
%
%
%
%
%
\begin{teo}\label{thm:Bhat}
Under the assumption that $\bm{\xi^m}_l  \sim N_p(\langle\bm B_l, \bm{X^O}\rangle, {\Sigma}_l^{m})$, the optimal linear functional $\bm{\widehat B}_l$, defined as the solution of the problem (\ref{eqn:Prob_Bhat}), satisfies the following matrix equation:
\begin{equation} \label{sistema}
\mC^{\bm{OO}} \bm{\widehat{B}}_l = \bm{R}_l,
\end{equation}
where $\mC^{\bm{OO}}$ is the covariance operator of the process $\bm{X^O}$ with kernel $\bm{C^{OO}} = \left\{C_{hk}^{O_hO_k}\right\}$, and  
\[
\mC^{\bm{OO}} \bm{\widehat{B}}_l = (\mC^{\bm{OO}}\bm{\hat\beta}_{1l}\mid\cdots\mid \mC^{\bm{OO}}\bm{\hat\beta}_{pl}), 
\quad
%
\bm{R}_l =  
\begin{pmatrix}
\mC_{11}^{O_1M_1} \varphi^{M_1}_l & \dots & \mC_{1p}^{O_1M_p} \varphi^{M_p}_l \\
\vdots & & \vdots \\
\mC_{p1}^{O_pM_1} \varphi^{M_1}_l & \dots & \mC_{pp}^{O_pM_p} \varphi^{M_p}_l
\end{pmatrix}.
\]
\end{teo}
The proof is reported in Appendix~\ref{appendixA}. We notice that the system~(\ref{sistema}) is a multivariate generalization of \cite{Kraus} and reveals that $\bm{\widehat{B}}_l$ does not depend on the precision matrix $\Theta^m_l$. Consequently the problems of estimating $\bm{B}_l$ and $\Theta_l$ can be solved separately. However, the covariance operator $\mC^{\bm{OO}}$ is compact with an infinite dimensional range, which implies that the inverse operator $(\mC^{\bm{OO}})^{-1}$ is unbounded and therefore a small perturbation in $\bm{R}_l$ may lead to a perturbation of $\bm{\widehat B}_l$. In other words, as in the univariate setting, the inverse problem~(\ref{sistema}) is ill-posed. 

To solve this problem,  we modify the original ill-posed inverse problem in such a way that it becomes a well-posed problem with a stable solution. Similarly to~\cite{Kraus}, we propose to use a ridge regularization so that, instead of problem~(\ref{sistema}) we solve the problem
\begin{equation}\label{eqn:hatB_a}
\mC^{\bm{OO}}_\alpha \bm{\widehat{B}}^\alpha_l = \bm{R}_l,
\end{equation}
where $\mC^{\bm{OO}}_\alpha = \mC^{\bm{OO}} + \alpha \mathcal{I}^{\bm O}$, with $\alpha$ a positive tuning parameter and $\mathcal{I}^{\bm O}$  the identity operator. Since the inverse of the operator $\mC^{\bm{OO}}_\alpha$ is bounded, the solution $\bm{\widehat{B}}^\alpha_l = \left(\mC^{\bm{OO}} + \alpha \mathcal{I}^{\bm O}\right)^{-1}\bm{R}_l$ is stable. In particular, the stability of the solution increases with $\alpha$, though at the expense of an increased bias as the system~(\ref{eqn:hatB_a}) deviates more and more from the original problem. As too small values of $\alpha$ are not sufficient to regularize the solution, a trade-off between bias and variance is needed. This aspect  will be discussed in detail in the next section.


\subsection{Computational aspects\label{sec:CompAsp}}

Suppose that we have a set of $n$ independent and identically distributed realizations from the graphical model $\{MGP(\mC), \mG\}$, where the covariance operator $\mC$ is partially separable with basis $\{\varphi_l\}_{l = 1}^{+\infty}$. We also assume that a subset of the $n$ realizations, denoted by $\mathbb{O}\subset\{1,\ldots,n\}$, is completely observed. \bt{Letting then $\overline{\mathbb{O}}$ be the complement of the set $\mathbb{O}$, we have that}, for each $i\in \overline{\mathbb{O}}$, the $j$th element of the random vector $\bm X_i$ is observed in the subset $O_{ij}$, while it is missing in $M_{ij}$. To emphasize the dependence of $\bm X_i$ from these subsets, the observed and missing fragments of $X_{ij}$ are denoted by $X_{ij}^{O_{ij}}$ and $X_{ij}^{M_{ij}}$, respectively.

As described in \bt{Section~\ref{sec:EStep}}, the preliminary step of the proposed procedure is to estimate the eigenfunctions $\varphi_l$ and the number of terms used to approximate the multivariate Karhunen-Loève expansion~(\ref{eqn:MKL}). To solve these problems, we follow~\cite{Kraus} and estimate the covariance operator by estimating the covariance functions as follows
\[
\widehat{C}_{hk}(s,t) = \frac{I_{hk}(s,t)}{\sum_{i=1}^{n}U_i^{hk}(s,t)}\sum_{i=1}^{n} U_i^{hk}(s,t) \left\{X_{ih}(s) - \hat{\mu}_{h}(s, t) \right\} \left\{X_{il}(t) - \hat{\mu}_{k}(s, t) \right\},
\]
where $U_i^{hk}(s,t)$ is equal to $1$ if $s\in O_{ih}$ and $t\in O_{ik}$ while it is $0$ otherwise, and $I_{hk}(s,t)$ is the indicator $\mathbbm 1_{\left\{\sum_{i=1}^{n} U_i^{hk}(s,t) > 0\right\}}$, while the mean functions are estimated by
\[
\hat{\mu}_{h}(s, t) = \frac{I_{hk}(s,t)}{\sum_{i=1}^{n}U_i^{hk}(s,t)}\sum_{i=1}^{n} U_i^{hk}(s,t) X_{ih}(s).
\]
Given the estimated covariance functions $\widehat{C}_{hh}$, one obtains the estimated covariance operators $\widehat{\mC}_{hh}$ and, consequently, the operator $\widehat{\mH} = p^{-1} \sum_{h = 1}^p \widehat{\mC}_{hh}$. From this the eigenfunctions are estimated by solving the empirical counterpart of the identity in Step~1 of the proposed pseudo-code, i.e.,
\[
\langle\widehat{\mH}(s,\cdot),\hat\varphi_l\rangle = \hat\lambda_l\hat\varphi_l(s).
\]
\bt{As mentioned already in section~\ref{sec:EStep}, the estimator $\widehat{\mH}$ obtained in this way   uses the information contained in the entire functional dataset and therefore provides an optimal initial estimate of the eigenfunctions, with no need for further recalculation at the next steps of the algorithm}.
Then, the number of $L$ terms used to approximate expansion~(\ref{eqn:MKL}) is chosen as the minimum number of components that explain a fixed proportion of variance explained \bt{(PVE). Differently to the more common use of this threshold in FPCA analysis, for the purpose of the methodology proposed in this paper, this value should be set as large as possible, so as to obtain the best approximation of the curves without encountering numerical values with eigenvalues that are too close to zero.}

Given the estimated eigenfunctions $\{\hat\varphi_l\}_{l = 1}^L$, for each $i\in \overline{\mathbb{O}}$, we impute the missing score $\bm\xi_{il}^{\bm {m}}$ using $\bm{\widehat\mu}^{m\mid o}_{il} = \langle \bm{\widehat B}^{\hat\alpha}_{il}, \bm {X^{O_i}}_i\rangle$, where $\bm{\widehat{B}}^{\hat\alpha}_{il} = \left(\mC^{\bm{O_iO_i}} + \hat{\alpha} \mathcal{I}^{\bm O_i}\right)^{-1}\bm{R}_{il}$ and $\hat{\alpha}$ is the value of the regularization parameter $\alpha$ that provides the optimal reconstruction of the full set of partially observed functions. As in~\cite{Kraus}, we consider the minimization problem:
\[
\min_{\mathcal A\,:\|\mathcal A\|_\infty < \infty} \mathbb{E} \| \bm X^{\bm M} - \mathcal A\bm X^{\bm O}\|^2,
\]
where the solution is in the class of continuous (bounded) linear operators from $L^2_{\bm O}$ to $L^2_{\bm M}$. By Fréchet differentiation, we obtain the system of normal equation $\mathcal A\mathcal C^{\bm{OO}} = \mathcal C^{\bm{MO}}$. Then, analogously to the multivariate principal scores, the solution is $\hat{\mathcal A}^\alpha = \mathcal C^{\bm{MO}}(\mathcal C^{\bm{OO}} + \alpha \mathcal I^{\bm{OO}})^{-1}$ and the optimal reconstruction of $\bm X^{\bm M}$ is $\hat{\mathcal A}^{\alpha} \bm X^{\bm O}$. From a practical point of view, the optimal $\alpha$-value is selected as the minimizer of the following generalized cross-validation criterion:
\begin{equation}\label{eqn:gcv}
\mathrm{gcv}_{i}(\alpha) = \frac{\sum_{i'\in\mathbb{O}} \|\bm X_i^{\bm M_i} - \hat{\mathcal A}^{\alpha} \bm X_i^{\bm O_i}\|^2}{\{1  - \mathrm{df}_{i}(\alpha)/|\mathbb{O}|\}^2},
\end{equation}
where $\mathrm{df}_{i}(\alpha)$ is the number of effective degrees of freedom, formally defined as:
\[
\mathrm{df}_i(\alpha) = \mathrm{tr}\{(\widehat{\mC}^{\bm{O_iO_i}}_\alpha)^{-1}\widehat{\mC}^{\bm{O_iO_i}}\} = \sum_{k = 1}^{+\infty}\frac{\hat\lambda_{k}^{\bm{O_iO_i}}}{\hat\lambda^{\bm{O_iO_i}}_{k} + \alpha}.
\]
Then, the missing scores are estimated using $\bm{\widehat\mu}^{m\mid o}_{il} = \langle \bm{\widehat B}^{\hat\alpha}_{il}, \bm {X^{O_i}}_i\rangle$, and $\bm{\hat\xi}_l$ are computed as described in Step~6 of the pseudo-code, i.e., by letting $\bm{\hat\xi}_{il} = \bm{\hat\xi^{o}}_{il} + \bm{\widehat \mu^{m \mid o}}_{il}$, where
\[
\hat\xi_{ijl}^o = \langle X^{O_{ij}}_{ij} - \hat\mu_j^{O_{ij}}, \hat{\varphi}^{O_{ij}}_l\rangle, \quad\text{and}\quad \hat\mu_{ij}(t) = \frac{J_j(t)}{\sum_{i = 1}^n O_{ij}(t)} \sum_{i = 1}^n O_{ij}(t) X_{ij}(t),
\]
and $O_{ij}(t)$ is used for the indicator $\mathbbm 1_{\{t\in O_{ij}\}}$, while $J_j(t) = \mathbbm 1_{\{\sum_{i = 1}^n O_{ij}(t) > 0\}}$.

Once the imputation of missing scores is complete, the precision matrices are estimated as described in Step~8 of the proposed algorithm, that is, by maximizing the following function:
\[
\sum_{l = 1}^{L}\left\{\log\det \Theta_l - \mathrm{tr}(S_l\Theta_l)\right\} - P_{\bm\gamma}(\{\Theta\}),
\]
where $S_l = \sum_{i = 1}^n \bm{\hat\xi}_{il}\bm{\hat\xi}_{il}^\top / n$ denotes the sample covariance matrix. The previous steps are repeated until a suitable convergence criterion is met and the edge set $\mE$ is estimated by $\widehat\mE = \bigcup_{l=1}^{L}\widehat\mE_l$, where $\widehat\mE_l = \{(h,k)\;:\; \hat\theta_{hkl}\ne0\}$.

\subsection{Selection of tuning parameters} \label{sec:tuning}
Different values of the tuning parameters $\gamma_1$ and $\gamma_2$ correspond to different levels of sparsity in the associated graph. We select the optimal values by minimizing the extended Bayesian Information Criterion (eBIC) \citep{EBIC}. As the method we propose follows an EM-type algorithm, we consider the approach proposed by \cite{Ibrahim2008} and replace the marginal likelihood by the Q-function evaluated at the optimum
\[
Q(\{\widehat{\Theta}^{\gamma_1,\gamma_2} \}) = \sum_{l = 1}^{L}\left\{\log\det \widehat\Theta_l^{\gamma_1,\gamma_2} - \mathrm{tr}(S_l^{\gamma_1,\gamma_2}\widehat\Theta_l^{\gamma_1,\gamma_2})\right\},
\]
which is readily available at convergence. Here
 $\{\hat{\Theta}^{\gamma_1,\gamma_2} \}$ are the precision matrices that maximize the penalized loglikelihood function, where we now explicitly emphasize the dependence on the tuning parameters $\gamma_1$ and $\gamma_2$. Therefore, similarly to~\cite{SottileEtAl_JRSSC_24}, the eBIC criterion that we use is given by
\[
eBIC(\gamma_1,\gamma_2)= -2Q(\{\widehat{\Theta}^{\gamma_1,\gamma_2} \}) + \sum_{l = 1}^L\left\{\vert \widehat\mE^{\gamma_1,\gamma_2}_l \vert \log n + 4\tau_1 \vert \widehat\mE^{\gamma_1,\gamma_2}_l \vert \log p\right\}
\]
where $\widehat\mE^{\gamma_1,\gamma_2}_l$ denotes the number of edges in the graph associated to $\widehat{\Theta}^{\gamma_1,\gamma_2}_l$ and $\tau_1$ is set to 0.5. 

\section{Simulation study} \label{sec:simulation}
This section investigates the finite-sample performance of the proposed methodology \bt{by an extensive simulation study specifically designed to evaluate important aspects of the methodology, namely curve reconstruction accuracy, precision-matrix estimation, graph recovery, and computational cost. We also include a comparison with existing procedures \citep{Kraus} and a study of sensitivity to empirical choices that are made in the inferential procedure.} The data-generating mechanism is described in Section~\ref{sec:setting}, together with the evaluation criteria. Section~\ref{sec:evaluation} presents the results of the simulations under a broad collection of scenarios. 
%
%

\subsection{Simulation design, competitors and evaluation criteria\label{sec:setting}} 
\bt{We use the following procedure to simulate partially observed functional data from a FGGM, under the assumption of partial separability of the covariance operator. Each random function is modelled as $\bm X_i(t) = \sum_{l = 1}^L\bm\xi_{il}\varphi_l(t)$, with $i = 1,\ldots, n$ and $\{\varphi_l\}_{l=1}^L$ is a Fourier basis system. In this study, the sample size is fixed to $n = 100$. In order to generate a discretized version of the random functions, each basis is evaluated on an equally spaced grid of $50$ time points, i.e., $t_1 < \cdots <t_{50}$, with $t_1 = 0$ and $t_{50} = 1$. The random vectors $\bm\xi_{il}$ are  drawn independently from a multivariate Gaussian distribution with mean zero and covariance matrix $\Sigma_l = a_l\Theta_l^{-1}$, where $a_l$ is a decay factor ensuring that $\tr{\Theta_l}$ decreases monotonically in $l$. As in~\cite{ZapataEtAl_BioK_22}, we let $a_l = 3l^{-1.8}$. To simulate a sparse precision matrix we use the \texttt{bdgraph.sim} function, available in the \texttt{BDgraph} R package~\citep{MohammadiEtAl_JSS_19}, with a small-world graph structure and a number of links equal to 10\% of the maximum number of links, i.e., $p \times (p - 1) / 2$. In this study, all the precision matrices share the same sparse structure, i.e., $\theta_{hkl} \ne 0$ for each $l = 1,\ldots, L$.}

\bt{The number of partially observed random functions, i.e., the cardinality of the set $\overline{\mathbb{O}}$, is related to the sample size by the identity $\mathrm{card}{(\overline{\mathbb{O}})} = \pi_{po}n$, where $\pi_{po}$ denotes the proportion of partially observed data. The set $\overline{\mathbb{O}}$ is constructed by randomly drawing $\pi_{po}n$ indices from the set $\{1,\ldots,n\}$. Then, for each $i \in \overline{\mathbb{O}}$ and each $h\in\{1,\ldots,p\}$, we simulate the random interval $M_{ih} = [a_{ih}, b_{ih}]$, where $a_{ih}$ is drawn from a uniform distribution over the interval $[0, 1 - \pi_w]$, $b_{ih} = a_{ih} + \pi_w$. Consequently, $\pi_w$ is the length of the random interval. Finally, $X_{ih}(t_k)$ is missing if $t_k\in M_{ih}$. In the following sections, we evaluate the effects of changing the parameters defined earlier. Specifically, we use the scenario with $L = 5$, $p = 20$, $\pi_{po} = 0.5$ and $\pi_w = 0.5$ as the baseline scenario.}

\bt{We compare the proposed approach, denoted by \texttt{poFGGM}, with two natural alternatives: (i) a baseline estimator in the absence of missing data, denoted by \texttt{oracle}, and (ii) an approach in which the partially observed curves are first completed through imputation and subsequently used for graph inference. In the latter approach, we use the univariate method of \citet{Kraus} to complete each curve and therefore denote it by \texttt{Kraus}. Under this approach, each component of the multivariate Gaussian process is imputed independently, thereby disregarding cross-sectional dependences. To ensure a fair comparison, we use the same FPCA-derived basis as in our proposed estimation procedure, so that any differences in performance can be attributed exclusively to the estimation step rather than to discrepancies in the basis expansion. After imputation, the joint graphical lasso estimation procedure is applied to the completed data to estimate the precision matrices, as in the M-step of the proposed method. For the proposed approach, the estimation of the precision matrices (Step~8 of Algorithm~\ref{tbl:PseudoCode}) is carried out using the algorithm of \citet{SottileEtAl_JRSSC_24}. The procedure is applied over a grid consisting of 21 equally spaced values of $\gamma_1$ between $0$ and $\gamma^{\max}_{1}$, where $\gamma^{\max}_{1}$ denotes the smallest value of $\gamma_1$ for which all off-diagonal elements of the estimated precision matrices are equal to zero. The tuning parameter $\gamma_2$ is fixed to zero since all precision matrices share the same sparse structure. Finally, for all methods under consideration, the number of basis functions is selected by applying FPCA to the partially observed data and retaining enough components to explain at least 99\% of the variance (PVE = 0.99). As discussed earlier, this large threshold was chosen to avoid potential reconstruction errors arising from the use of an insufficient number of basis functions.}

\bt{The performance of the  methods is assessed by a number of criteria, which are computed using, for each scenario, 100 Monte Carlo samples. Specifically, the accuracy of the precision matrix estimation is measured through the mean relative Frobenius error:
\[
\mathrm{Err}_\Theta = \frac{1}{L}\sum_{l = 1}^L
\frac{\|\widehat\Theta_l-\Theta_l\|_F}{\|\Theta_l\|_F},
\]
whereas the graph recovery property is evaluated by the area under the ROC curve (AUC), where each point of the curve corresponds to the sensitivity and specificity of edge selection at a specific value of the tuning parameter $\gamma_1$. The performance in terms of curve reconstruction is evaluated through the mean relative reconstruction error:
\[
\mathrm{Err}_X = \frac{1}{p\,\mathrm{card}{(\overline{\mathbb{O}})}}\sum_{i \in \overline{\mathbb{O}}}\sum_{h = 1}^{p} \frac{\|\widehat X_{ih} - X_{ih}\|^2}{\|X_{ih}\|^2}.
\]
Finally, the computational cost of the proposed procedure is evaluated through the wall-clock elapsed time.}

\subsection{Evaluation and comparison of the proposed methodology}\label{sec:evaluation}
\bt{We start the simulation study with a comparison of the proposed method with existing approaches. We then proceed with a sensitivity analysis, where we consider the baseline configuration described above and vary one parameter at a time while keeping all others fixed at the values specified above.}

\bt{\textbf{How the performance compares with alternative approaches.} In this section, we focus on a direct comparison of the proposed method with the competitors described above under the same baseline setting scenario. Table~\ref{tab:n_p} reports the Monte Carlo mean and standard deviation of $\min_{\gamma_1}\mathrm{Err}_{\Theta}$, $\mathrm{AUC}_{\Theta}$, and $\min_{\gamma_1}\mathrm{Err}_{X}$. In the last two cases, we report the best results along the $\gamma_1$ path. Since the \texttt{oracle} method uses complete data, the curve reconstruction error is not reported in this case. The results of the simulation study show that the performance of the proposed method is in agreement with the \texttt{oracle} while it is significantly better than the univariate approach, particularly in terms of curves reconstruction. The same relative performance is observed for all the other simulation settings considered below.}
\bt{\begin{table}[b!]
	\caption{\bt{\small Simulation results comparing \texttt{poFGGM} with the univariate procedure of \texttt{Kraus} and the \texttt{oracle} estimator for the baseline scenario.}\label{tab:n_p}}
    \centering
	\resizebox{0.6\textwidth}{!}{%
		\centering
		\begin{tabular}[t]{cccccc}
			\toprule
		 & $\min_{\gamma_1}\mathrm{Err}_{\Theta}$ & & $\mathrm{AUC}_{\Theta}$ & & $\min_{\gamma_1}\mathrm{Err}_{X}$  \\ \cline{2-6}
		 \texttt{oracle} & 0.370 (0.015) & & 0.874 (0.037) & & --   \\
			 \texttt{poFGGM} & 0.370 (0.015) & & 0.875 (0.036) & & 0.019 (0.007)\\
              \texttt{Kraus} & 0.389 (0.014) & & 0.870 (0.035) & & 0.449 (0.041)  \\
			\bottomrule
		\end{tabular}%
	}
\end{table}}

\begin{figure}[t!]
	\centering
	\subfigure[Effects of increasing $p$ from $20$ to $120$, while keeping $L$ fixed to $5$.]{\includegraphics[width=\textwidth]{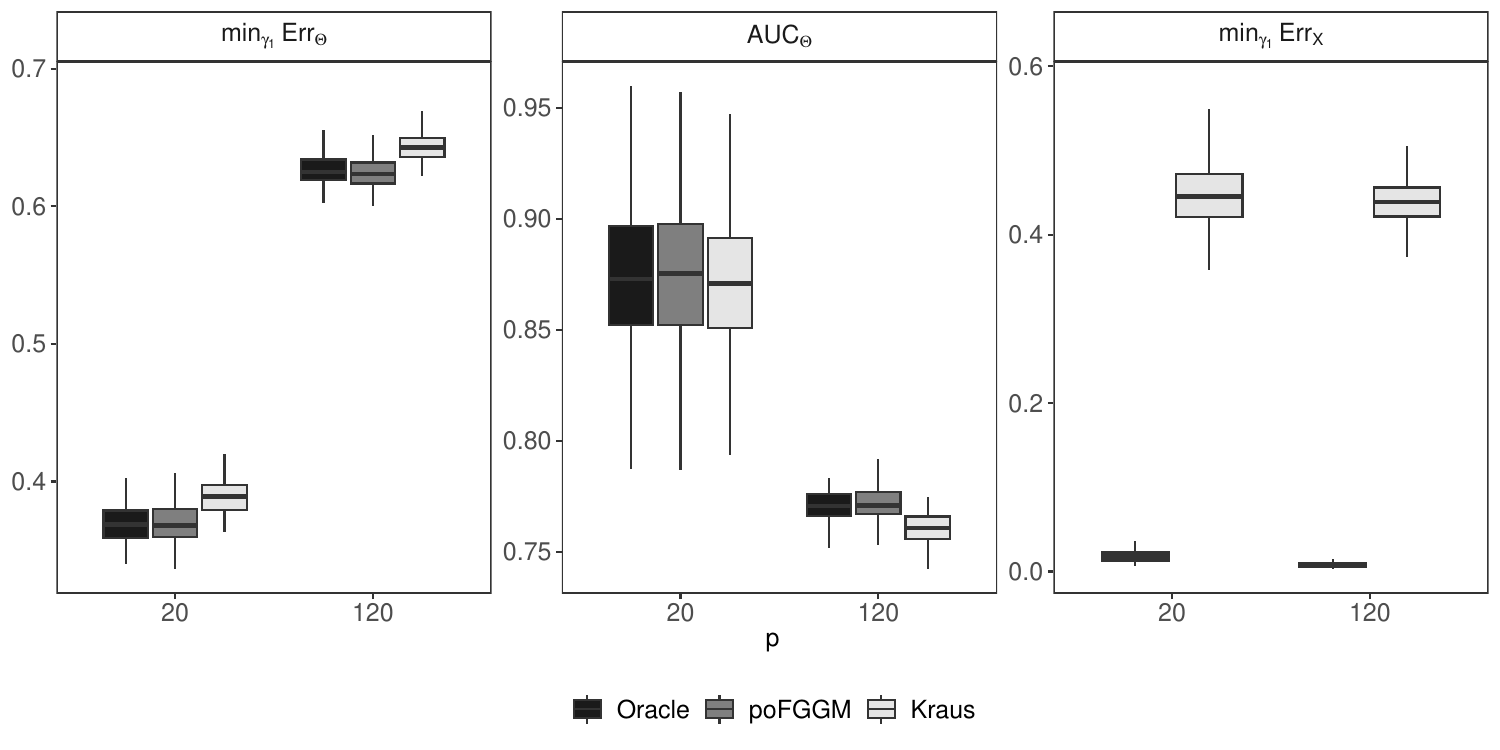}}
    \subfigure[Effects of increasing $L$ from $5$ to $9$, while keeping $p$ fixed to 20.]{\includegraphics[width=\textwidth]{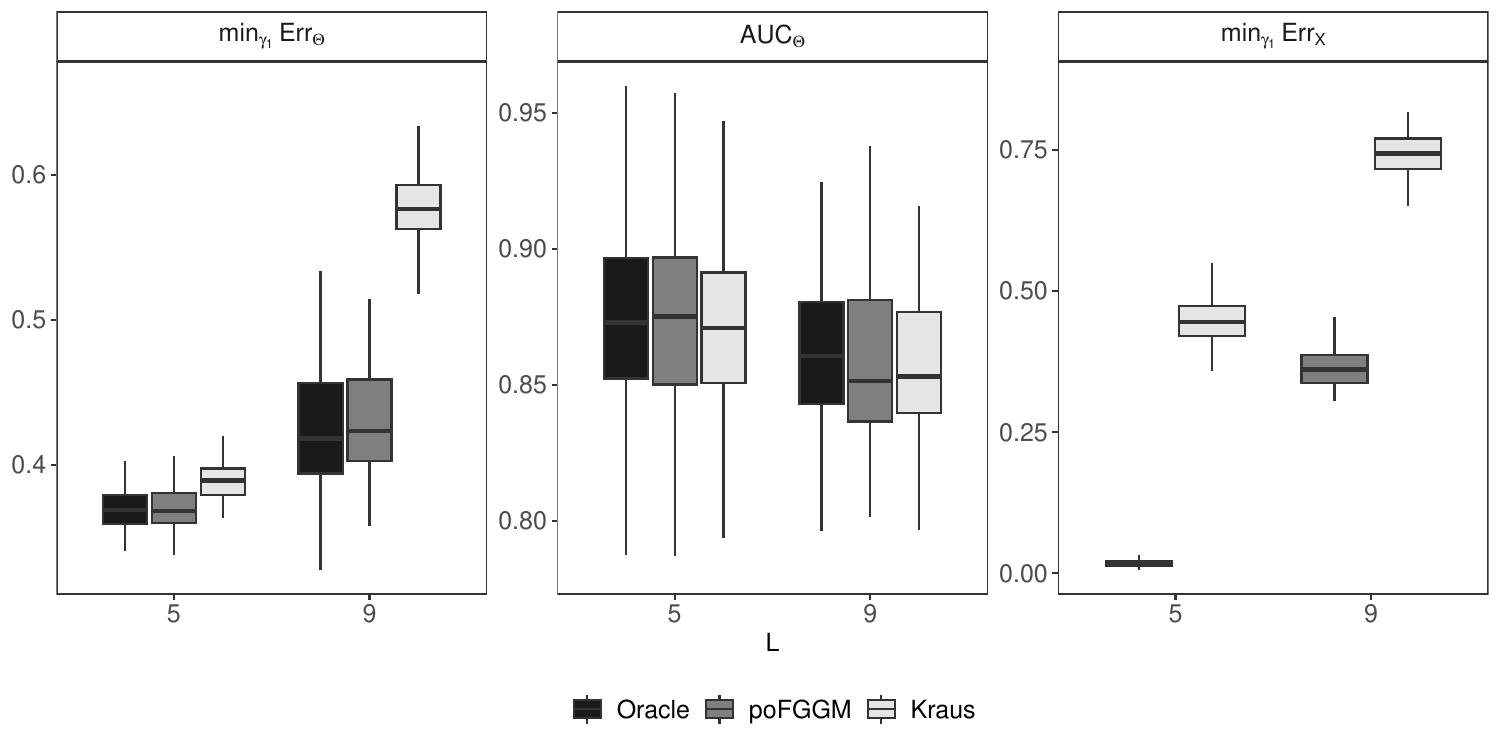}}
	\caption{\bt{\small Sensitivity analysis of the considered methods to the number of parameters.}\label{fig:p_l}}
\end{figure}

\bt{\textbf{How the performance varies with the number of parameters.} In this simulation, we evaluate the performance of the method as the number of parameters increases. This is achieved either by varying the number of variables, as this implies an increase in the dimensions of the precision matrices, or by varying the number of terms in the multivariate Karhunen–Loève expansion, as this corresponds to an increase in the number of precision matrices. In particular, we consider two specific scenarios. The first is obtained by keeping $L$ fixed at its baseline level, i.e., $L = 5$, and increasing $p$ from 20 to 120. The second is instead obtained by keeping $p$ fixed to 20 and increasing $L$ from $5$ to $9$. Figure~\ref{fig:p_l} shows how in both scenarios the performance generally deteriorates  as the number of parameters increases. However, the proposed method remains aligned with the oracle and is consistently more accurate than the univariate approach, particularly in terms of curve reconstruction.} 

\begin{figure}[t!]
	\centering
	\subfigure[Effects of varying $\pi_{po}$ while keeping $\pi_w$ equal to $0.5$.]{\includegraphics[width=\textwidth]{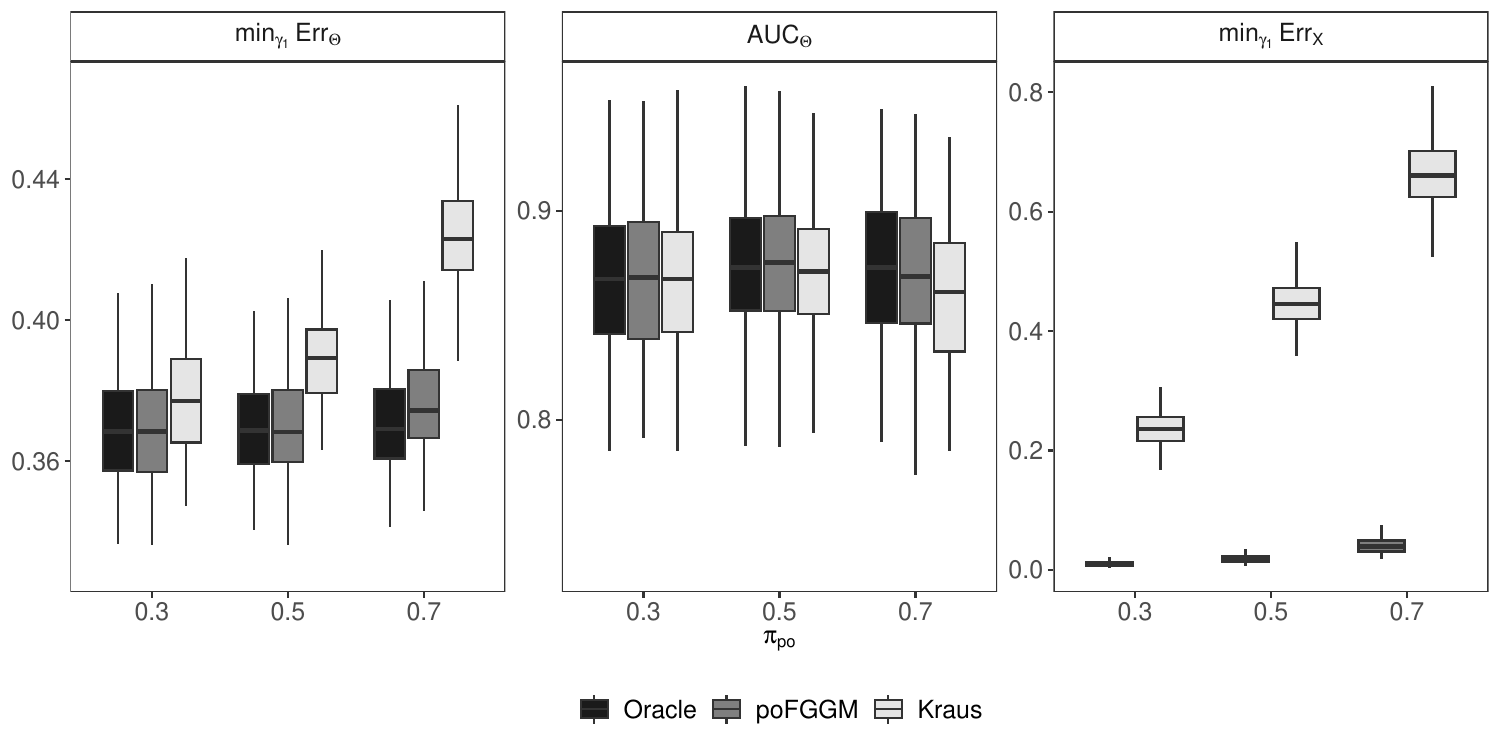}}
    \subfigure[Effects of varying $\pi_{w}$ while keeping $\pi_{po}$ equal to $0.5$.]{\includegraphics[width=\textwidth]{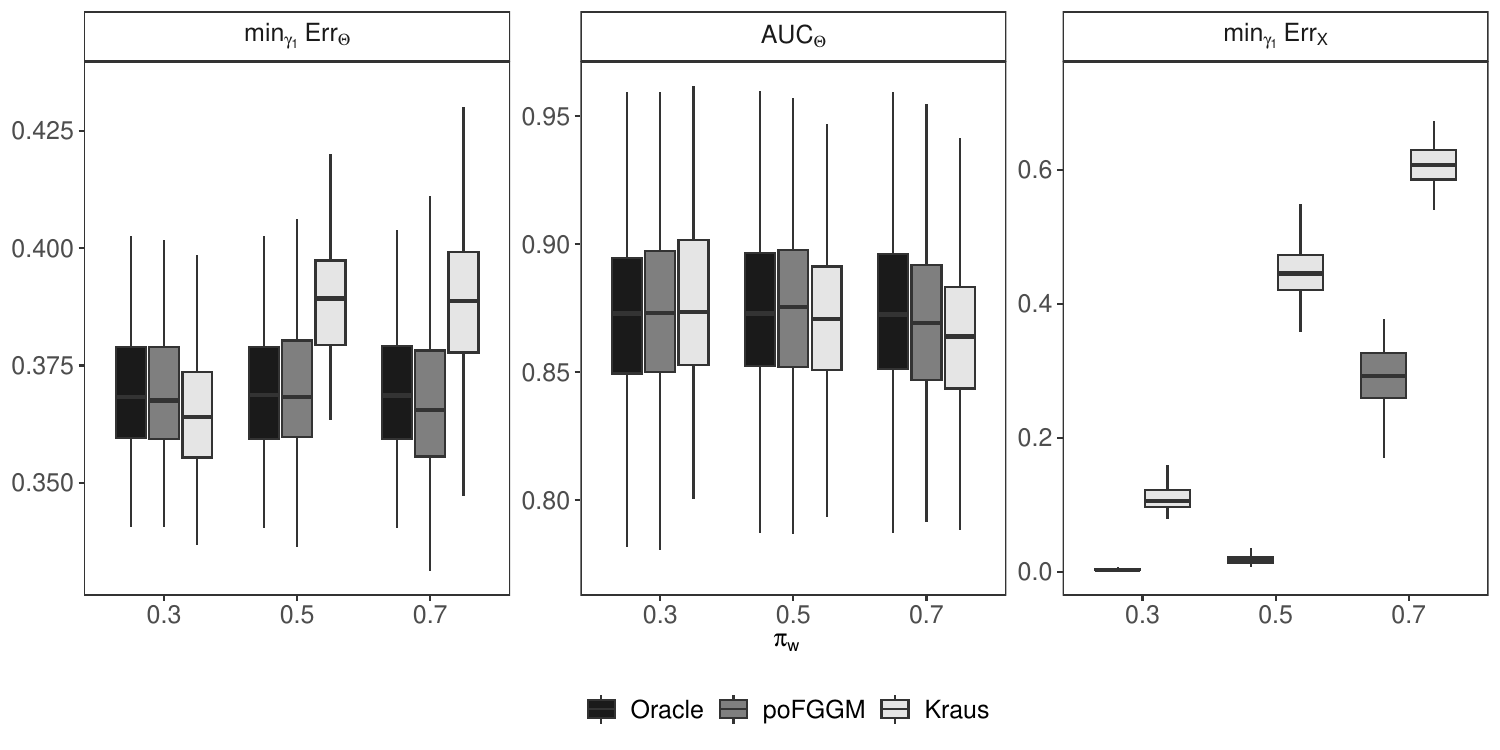}}
	\caption{\small Sensitivity analysis of the considered method to the level of missing data. \label{fig:missing_po}}
\end{figure}

\bt{\textbf{How the performance varies with the level of missing data.} We next study how the methods respond to increasing levels of missingness, which is controlled via the proportion of partially observed functional data, i.e., the variable $\pi_{po}$, and the length of the random intervals $M_{ih}$, i.e., the variable $\pi_w$. Again, we use two specific scenarios. The first is obtained by fixing $\pi_w = 0.5$ and varying $\pi_{po}$ over $\{0.3,0.5,0.7\}$, while the second is specified by varying $\pi_{w}$ over $\{0.3,0.5,0.7\}$ and keeping $\pi_{po} = 0.5$ fixed. All other parameters are set at the baseline level. Figures~\ref{fig:missing_po} shows how the proposed method is generally robust to the extent of missingness, while the performance of the univariate approach deteriorates significantly  as the percentage of missing increases.  As before, the proposed method consistently outperforms the univariate approach, both in terms of curve reconstruction and network inference, with the largest gains observed when the amount of missingness is the largest.} 
%
%

\begin{figure}[t!]
	\centering
	\includegraphics[width=\textwidth]{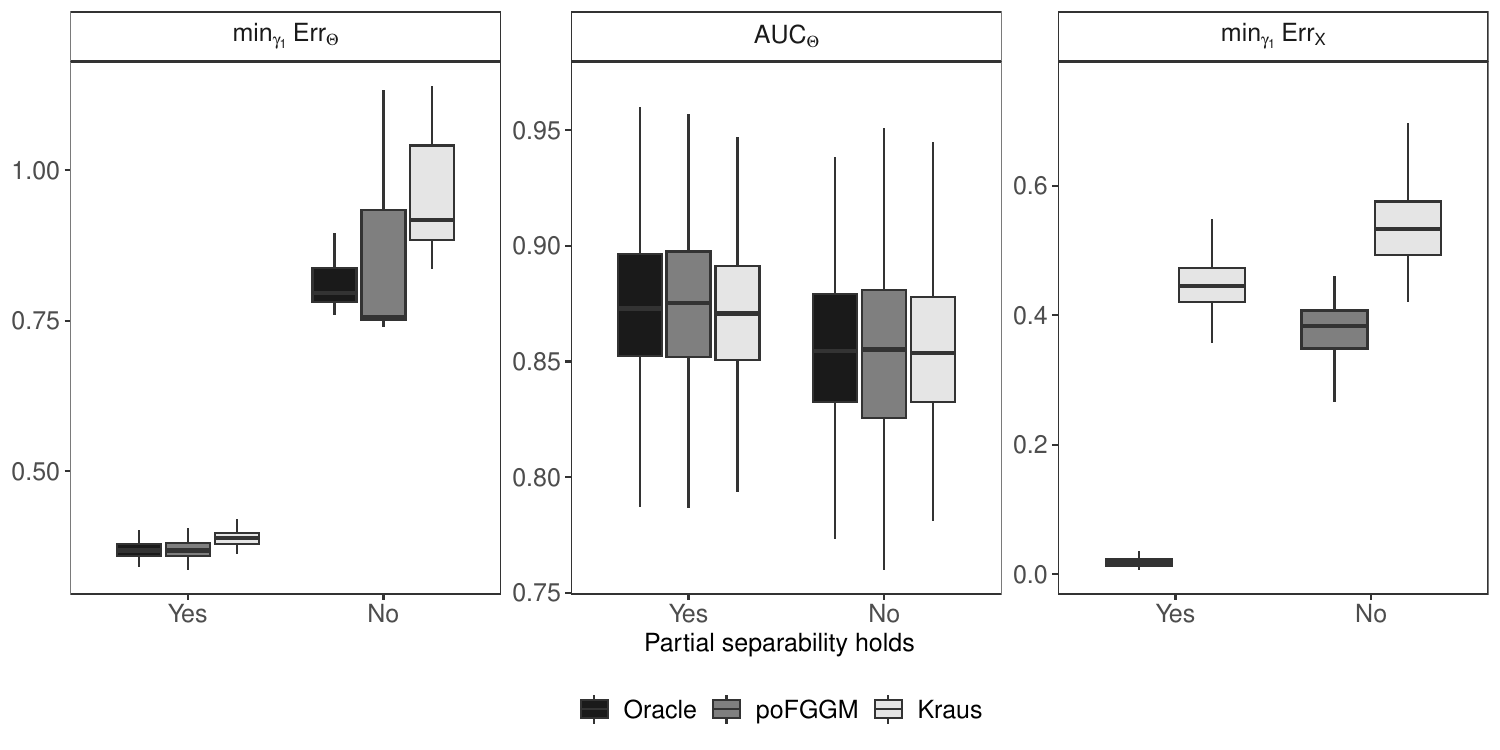}
	\caption{\small Sensitivity analysis of the considered methods to the assumption of partial separability. \label{fig:p_part_sep_sw}}
\end{figure}

\bt{\textbf{How the performance is affected by a violation of the assumption of partial separability.} In this simulation,  we consider a setting in which the assumption of partial separability is violated. To this end, we consider a non-partially separable data-generating mechanism by perturbing the baseline covariance structure using the procedure adopted by \citet{ZapataEtAl_BioK_22}. In particular, starting from covariance matrices generated under partial separability, we first construct a block-banded precision matrix $\Theta$ whose diagonal blocks are given by $\Theta_{l,l}=\Theta_l$, while the off-diagonal blocks satisfy
\(\Theta_{l+1,l}=\Theta_{l,l+1}=\rho\{\Theta_l^{*}+\Theta_{l+1}^{*}\}\) where \(\Theta_l^{*}=\Theta_l-\mathrm{diag}(\Theta_l)\).
The parameter $\rho=0.01$ is chosen so that the resulting full precision matrix is positive definite. The covariance matrix used in this misspecified setting is then defined as \(\Sigma_{\mathrm{nps}}=\mathrm{diag}(\Sigma)^{1/2}\Theta^{-1}\mathrm{diag}(\Sigma)^{1/2},\) which preserves the marginal scaling of the baseline model while inducing dependence across the components that is incompatible with partial separability.
The results in Figure~\ref{fig:p_part_sep_sw} indicate that the proposed method remains aligned with the oracle, i.e., to the case where the curves are fully observed, also in this scenario. Across the three performance measures, network recovery appears to be the most robust to model misspecification. 
This finding is consistent with the empirical evidence reported by \citet{ZapataEtAl_BioK_22}, who also observed that functional graphical model estimation based on partial separability can retain good network recovery performance under controlled violations from the assumption.}

%

\bt{\textbf{How the performance is affected by empirical choices.}} 
\bt{This last part of the simulation study assesses the sensitivity of the competing methods to user-specified parameters that are not directly related to the data generating process. Specifically, we investigate the selection of the number of empirical eigenfunctions for curve reconstruction, denoted by $\hat L$ and determined by the proportion of variance explained (PVE), and the choice of the ridge parameter $\alpha$.}

\begin{figure}[t!]
	\centering
	\subfigure[Sensitivity to the proportion of variance explained.]{\includegraphics[width=\textwidth]{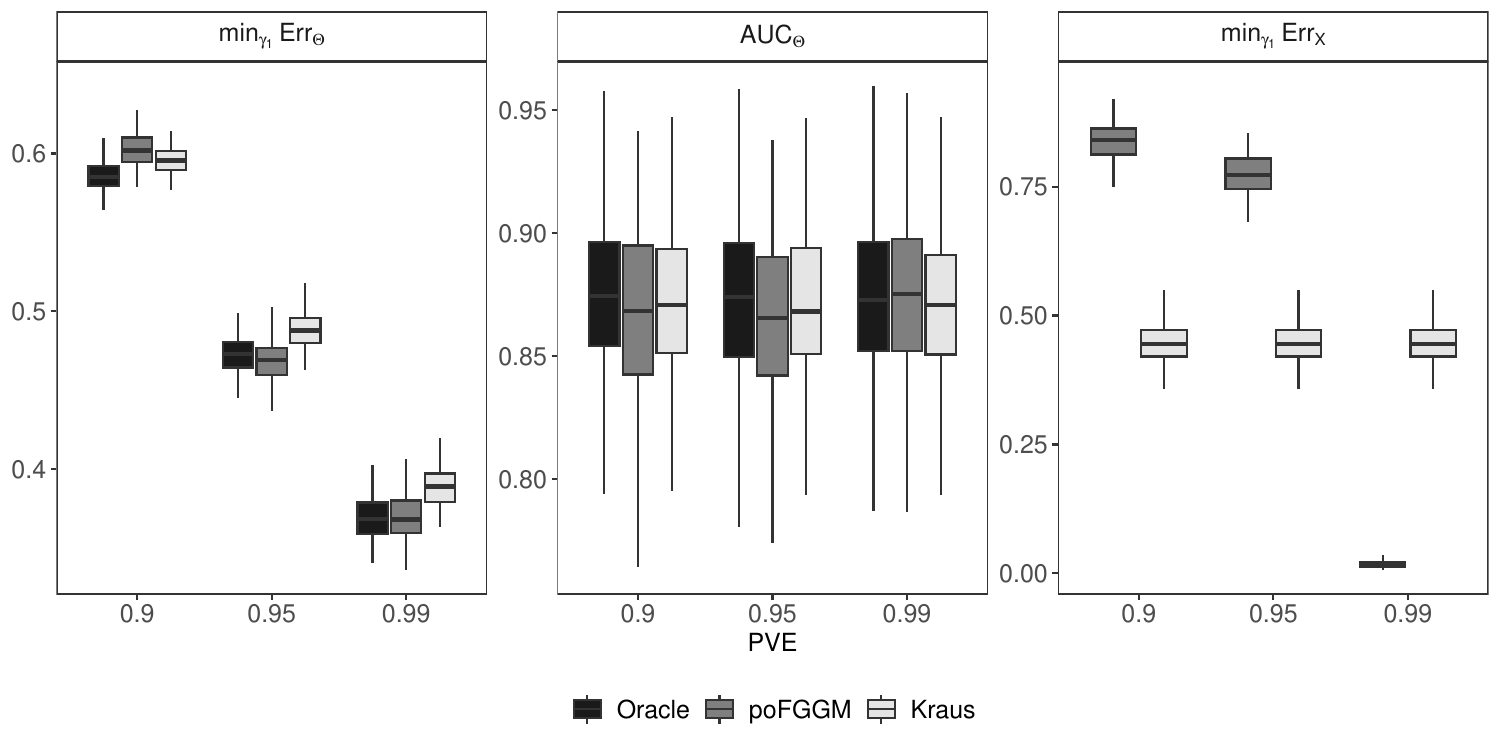}}
    \subfigure[Sensitivity to the selection of the ridge parameter.]{\includegraphics[width=\textwidth]{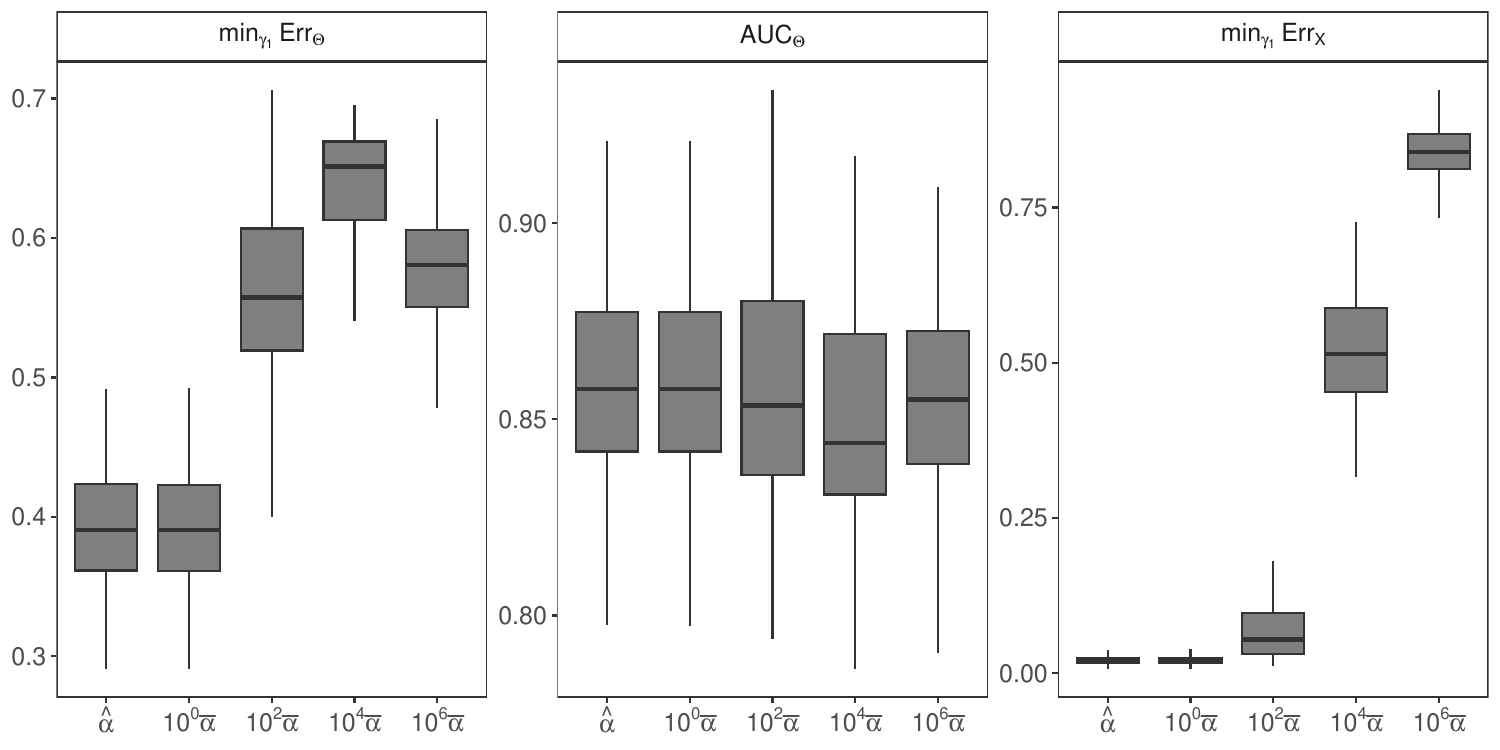}}
	\caption{\small Results of the simulation studies  evaluating the impact of the choice of the proportion of variance explained and the ridge parameter $\alpha$ on the performance of the proposed method. \label{fig:pev_alpha}}
\end{figure}

\bt{To assess the impact of $\hat L$, we lower the PVE baseline value of $0.99$ to  $0.90$ and $0.95$, respectively, while keeping all other parameters at their baseline values. Figure~\ref{fig:pev_alpha}a shows how the choice of this threshold is crucial  for curve reconstruction, with lower values leading to worse results than the univariate approach. This is then reflected also in a poorer performance in network inference, particularly in terms of estimation of the precision matrices.}

\bt{Next, we investigate the sensitivity of the proposed method to the choice of the ridge penalty parameter $\alpha$ used for reconstructing the partially observed functions. As described in Section~\ref{sec:CompAsp}, the proposed method employs the optimal value, denoted by $\hat{\alpha}$, obtained by minimizing equation~(\ref{eqn:gcv}) for each curve. To assess the impact of this choice, we compare it with a range of alternative values. Specifically, in each simulation run, we first compute the average optimal value across all curves, denoted by $\bar{\alpha}$. We then construct a sequence of candidate values by multiplying $\bar{\alpha}$ by $10^k$, with $k \in \{2,4,6\}$, thus generating increasingly larger departures from the data-driven choice of $\alpha$. The results in Figure~\ref{fig:pev_alpha}b show how the proposed method is robust to small departures from the optimal values while it starts to deteriorate as the distance increases.}

\bt{
\paragraph{Computational cost}
Table~\ref{supp:tab:comp_time} reports the computational time of the proposed method across all settings considered in the simulation study.  This is measured in terms of wall-clock elapsed time, reported in minutes for every scenario and method. Further details on the hardware and software configuration are provided in Appendix \ref{supp:comp_cost}. The table reports the mean and standard deviation across the $100$ replications, together with the factor change relative to the corresponding baseline configuration. The results show how the computational cost increases significantly  in the high-dimensional setting.}
\begin{table}[!htbp]
	\footnotesize
	\caption{\small Computational time (in minutes) for the proposed \texttt{poFGGM} method across different settings, reported as Monte Carlo mean and standard deviation across the $100$ iterations. In each case, we report the relative change compared to the baseline configuration ($n=100$, $p=20$, $L=5$, $\pi_w=0.5$, $\pi_{po}=0.5$).}
	\label{supp:tab:comp_time}
	\centering
	\resizebox{0.5\textwidth}{!}{%
		\begin{tabular}{lcc}
			\toprule
			  Scenario & Cost time & Relative to baseline \\
			\midrule
			Baseline & 0.364 (0.013) & -\\
			$\pi_{po}=0.3$ & 0.275 (0.019) & $0.8\times$ \\
			$\pi_{po}=0.7$ & 0.582 (0.034) & $1.6\times$ \\
			$p = 120$ & 20.082 (3.185) & $55.2\times$ \\
			$L=9$ & 0.721 (0.037) & $2.0\times$ \\
			\bottomrule
	\end{tabular}}
\end{table}

\bt{
\subsection{Summary of the simulation results}\label{sec:sim_summary}
The simulation study yields five main conclusions. First, the proposed method consistently outperforms the univariate approach of \citet{Kraus} in terms of curve reconstruction, across dimensional regimes and missingness configurations. In addition, its performance is often close to that of the \texttt{oracle} benchmark, further supporting the effectiveness of the proposed procedure. Second, the gains in curve reconstruction generally translate into better or comparable performance in precision matrix estimation and graph recovery. Third, the method remains effective in high-dimensional settings, although at a substantially larger computational cost. Fourth, the sensitivity analyses identify the tuning parameters that are most influential. In particular, both the PVE threshold and the ridge regularization parameter play a non-trivial role in curve reconstruction, and poor choices may also deteriorate precision-matrix estimation. 
Fifth, violations of the partial separability assumption lead to poorer performance in curve reconstruction and precision-matrix estimation, while they have  a more limited impact on network recovery.}

\bt{\section{Modelling dependences between ESG dimensions}
\label{sec:esg_application}}

\bt{We now illustrate the proposed  methodology on the sovereign ESG data introduced in Section~\ref{subsec:esg_motivating_example}. Formally, let $i=1,\dots,n$ indicate the countries and $j=1,\dots,p$ the ESG indicators. For each country $i$, indicator $j$, and calendar year $t$ in the analysis window, let $X_{ij}(t)$ denote the observed value of the indicator for that year. This is a natural setup for multivariate functional data, whereby  time plays the role of the functional domain, while each country is the statistical unit and is represented by a vector of interrelated indicator trajectories
\[
X_i(t)=\bigl(X_{i1}(t),\dots,X_{ip}(t)\bigr)^\top.
\]
The empirical goal is both to learn the joint dependence across ESG dimensions and to exploit this structure in the imputation of the partially observed curves.}

\bt{Some pre-processing is needed prior to the analysis. First of all, the indicators display markedly different scales and distributional shapes. Some variables are shares or percentages, whereas others are physical quantities, per-capita measures or macroeconomic variables. Direct application of FPCA or penalized precision-matrix estimation to the raw data would therefore be driven, at least in part, by differences in scale rather than by the underlying dependence structure. To mitigate this problem, we pre-process the data indicator by indicator. In particular, for those variables exhibiting strong right-skewness or large scale differences, we apply a logarithmic transformation prior to standardization. This choice stabilizes variability, reduces the influence of extreme values, and makes the temporal profiles of heterogeneous indicators more comparable in the subsequent functional analysis. We then center and scale the transformed observations cross-sectionally, year by year, as described in Section~\ref{sec:inference}. 
The resulting standardized trajectories are then used to estimate the common functional basis via FPCA. 
To this end, we retain $L=15$ components, which account for $99.99\%$ of the total variability. These components define the layers for the latent multivariate score representation on which the reconstruction and network estimation procedures are based.}

\bt{The model is estimated over a two-dimensional grid of $21\times 5$ values of the tuning parameter pair $(\gamma_1,\gamma_2)$, appropriately chosen so as to explore models of varying sparsity levels.
The final model is selected by minimizing the $eBIC$ criterion over this $(\gamma_1,\gamma_2)$ grid. As described in Section~\ref{sec:tuning}, the criterion is evaluated after maximum-likelihood refitting under the graph structure estimated at each $(\gamma_1,\gamma_2)$ pair. 
}
%
%
\begin{figure}[t!]
	\centering
\includegraphics[width=.8\textwidth]{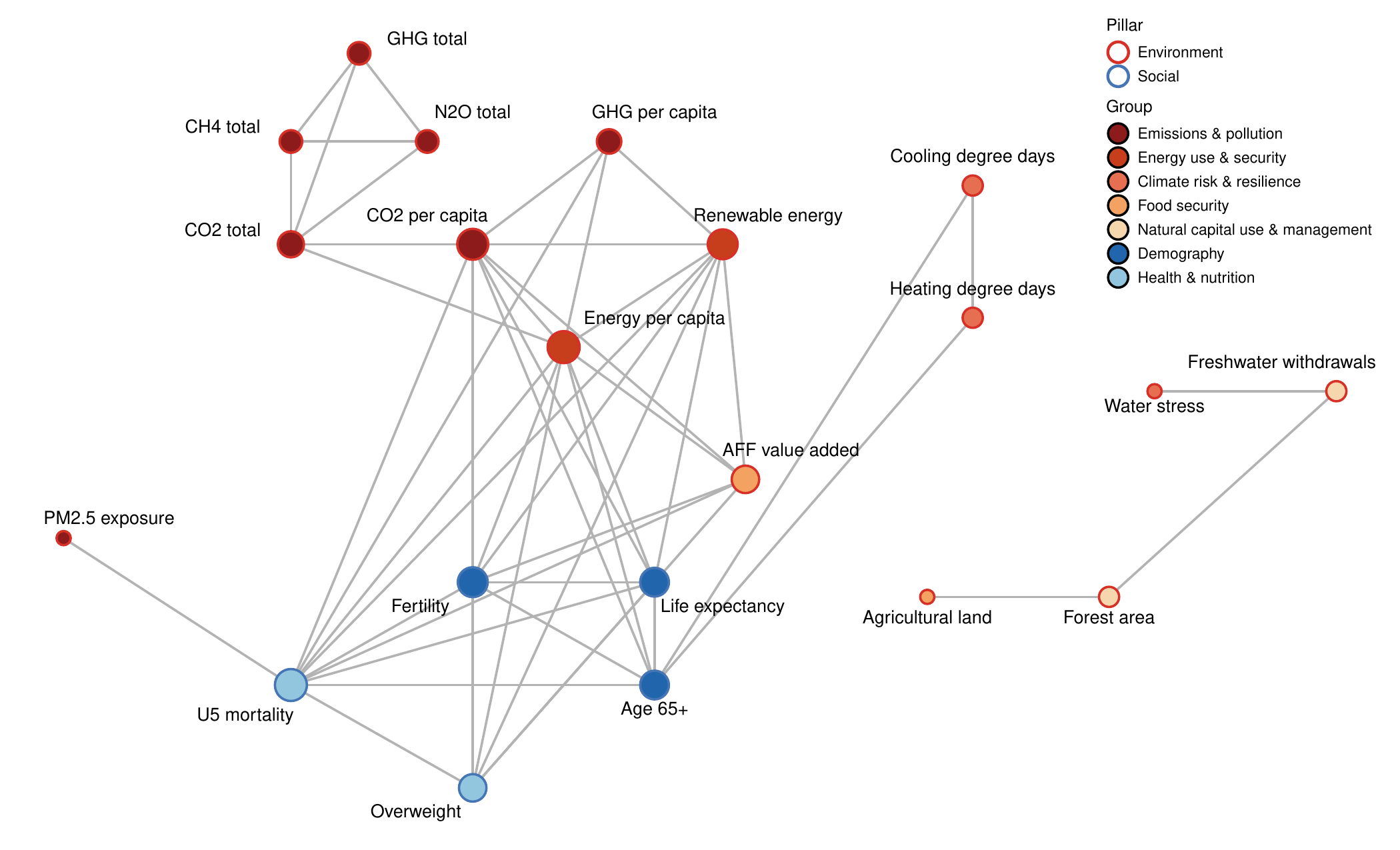}
	\caption{\small 
    Inferred functional dependence network among the sovereign ESG indicators. Node colors correspond to ESG pillars, while the shading within each pillar reflects the finer sub-groups reported in Table~\ref{tab:esg_indicators}. Isolated nodes are omitted for readability.}
	\label{fig:ebic_esg_net}
\end{figure}

\bt{Figure~\ref{fig:ebic_esg_net} shows the inferred network corresponding to the optimal model. Considering also the $12$ isolated nodes, which are not plotted to improve readability, the estimated graph is rather sparse, suggesting that the ESG indicators do not form a fully connected system, but rather organize themselves into a small number of inter-connected components. This is already useful information for more refined measures of aggregations of the indicators into scores. The interacting indicators belong to the E ad S pillars. We distinguish these nodes in Figure~\ref{fig:ebic_esg_net} with the use of red and blue colors, respectively. Looking at the sub-components of the graph, we see a natural association not only with the pillars, but also with their sub-blocks, as defined in Table~\ref{tab:esg_indicators} and visualized with a different shading of the two colors, respectively. However, the graph shows also a marked structure of dependence between the environmental and social indicators, suggesting that the system is governed by overlapping mechanisms that cut across these conventional categories. In particular, the graph shows a tight connection between the indicators related to gas emissions  and a group of social and demographic variables, namely fertility, life expectancy, the share of the population aged 65+ and under-five mortality. The CO$_2$ emissions per-capita indicator, one of the most connected nodes in the graph, appears to function as a bridge linking the aggregate emissions block to the other per-capita environmental indicators and the social outcomes. By contrast, the water indicators and land allocation indicators form an autonomous sub-network, separate from those related to emissions, energy and health.}

\begin{figure}[!htb]
	\centering
    \begin{minipage}{0.3\textwidth}
        \centering
        \includegraphics[width=\linewidth]{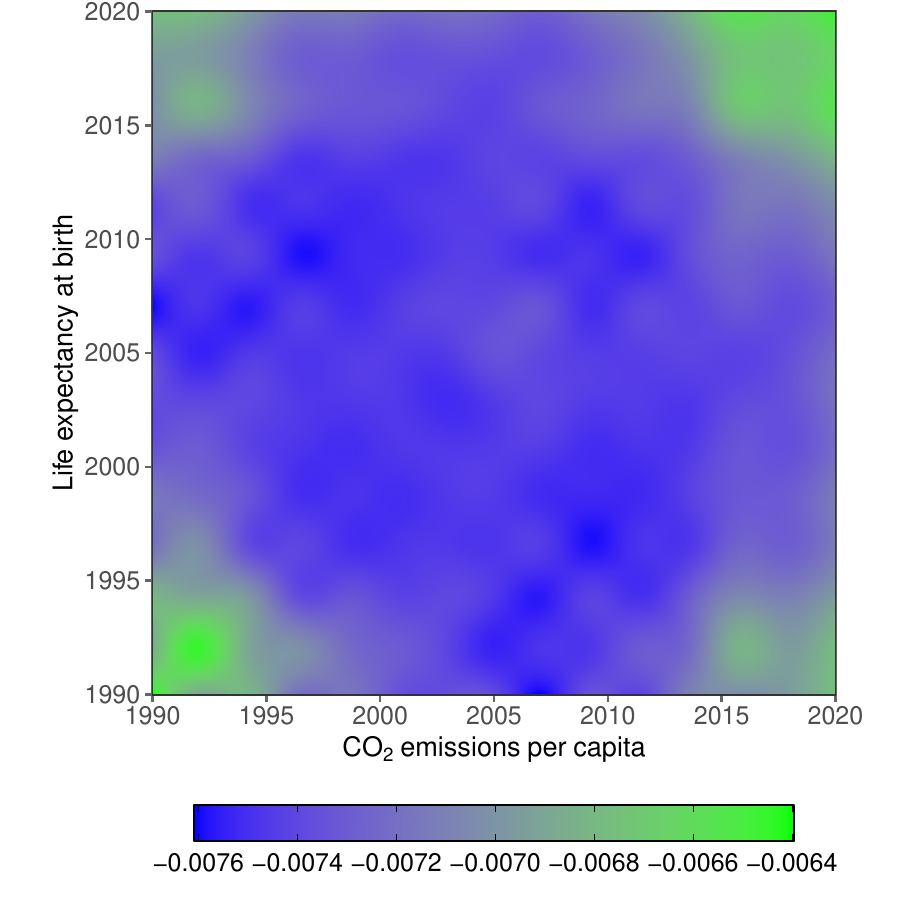}
    \centering (a)
    \end{minipage}
    \hfill
    \begin{minipage}{0.3\textwidth}
        \centering
        \includegraphics[width=\linewidth]{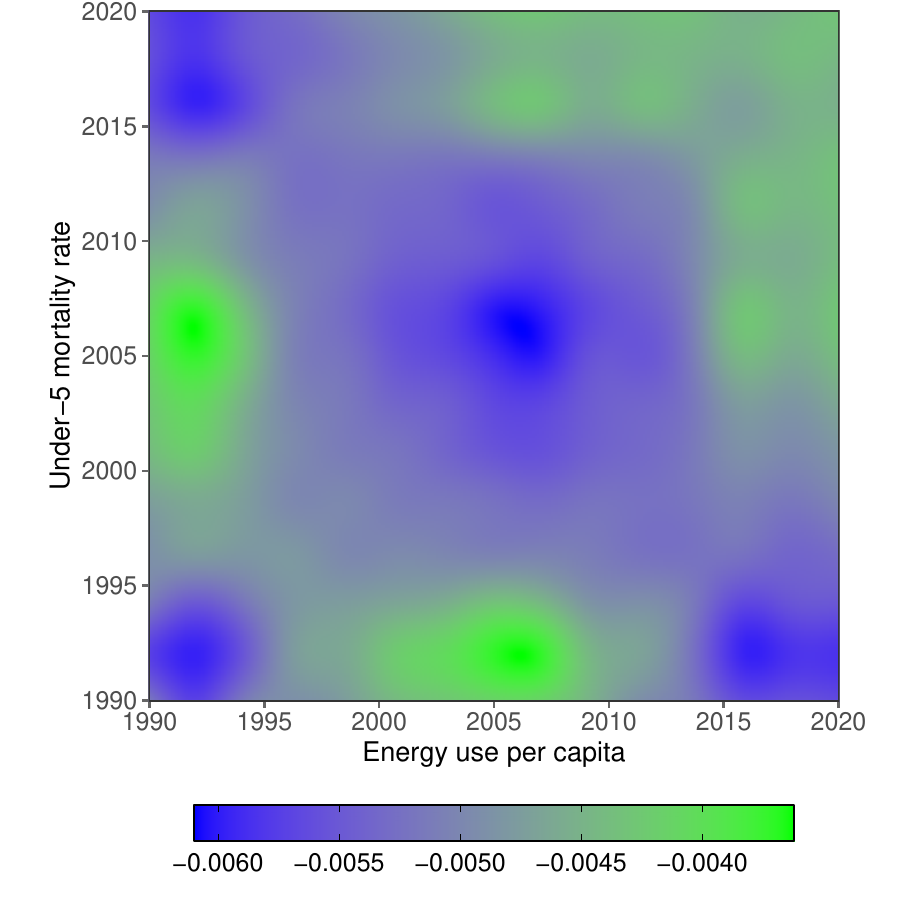}
    \centering (b)
    \end{minipage}
    \hfill
    \begin{minipage}{0.3\textwidth}
        \centering
        \includegraphics[width=\linewidth]{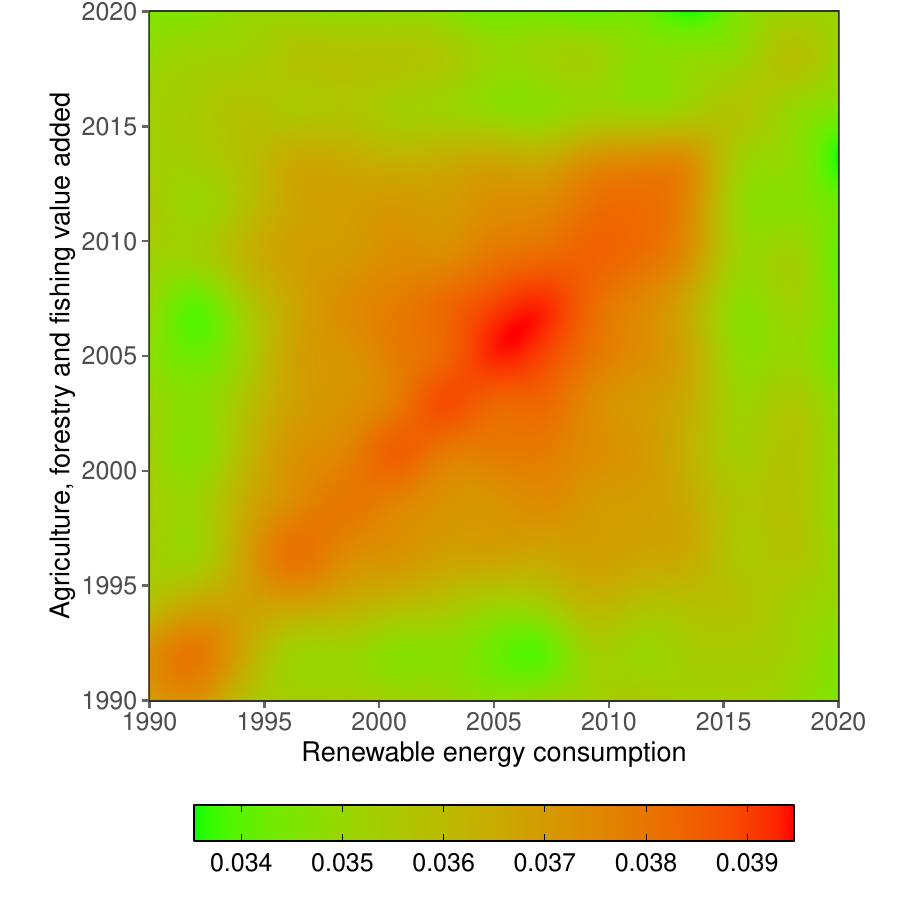}
    \centering (c)
    \end{minipage}
	\caption{\small Estimated cross-covariance surfaces for three inferred  relationships in the sovereign ESG illustration: (a) CO$_2$ emissions per capita versus life expectancy at birth; (b) energy use per capita versus the under-5 mortality rate; (c) renewable energy consumption versus agriculture, forestry and fishing value added.  The colors relate to the estimated cross-covariance at each pair of time points. }
	\label{fig:cross_cov}  
\end{figure}

\bt{To complement the analysis of the dependence structure, Figure~\ref{fig:cross_cov} investigates the sign and extent of the associations by plotting the estimated cross-covariance surfaces for three important relationships identified by the graph, namely CO$_2$ emissions per-capita versus life expectancy at birth (Figure~\ref{fig:cross_cov}a),  energy use per-capita versus the under-five mortality rate (Figure~\ref{fig:cross_cov}b), and renewable energy consumption versus agriculture, forestry and fishing value added (Figure~\ref{fig:cross_cov}c).}
\bt{Figures~\ref{fig:cross_cov}a and \ref{fig:cross_cov}b display predominantly negative surfaces, indicating persistent inverse co-movement between the corresponding pairs of indicators across countries and over time. In particular, Figure~\ref{fig:cross_cov}a suggests that higher per-capita CO$_2$ emissions are associated with lower life expectancy profiles, while Figure~\ref{fig:cross_cov}b indicates that countries with higher energy-use profiles tend to exhibit lower under-five mortality trajectories. In both cases, the association extends beyond contemporaneous time points, suggesting that the relation is not purely local in time. By contrast, Figure~\ref{fig:cross_cov}c shows a uniformly positive surface, indicating stable positive co-movement between renewable energy consumption and agriculture, forestry and fishing value added. The association is stronger around nearby time points and remains positive over the whole domain, pointing to a persistent link between energy mix and productive structure.}



\section{Conclusion}
We have presented a methodology for the reconstruction of partially observed multivariate curves and for the inference of the underlying functional Gaussian graphical model. The functional completion is done by the imputation of the missing part of the scores, which in turn is obtained through a functional linear regression where the predictors are represented by the observed part of the curves. This constitutes a generalization  of the procedure of \cite{Kraus} to the multivariate case. Under the assumption of partial separability introduced in \cite{ZapataEtAl_BioK_22}, and the Karhunen-Loève expansion of the multivariate functional data that results from that, the functional graphical model can be inferred using penalized likelihood approaches. Score imputation and graphical model fitting are integrated into an EM-type algorithm.

The simulation study evaluates the performance of the proposed method and shows how it is advantageous compared to existing approaches where score imputation is conducted for each variable separately, without accounting for the multivariate nature of the functional data. 
\bt{The illustration on sovereign ESG data further highlights the practical relevance of the proposed framework. In particular, the method is able to reconstruct fragmented country-level ESG trajectories and to uncover a sparse and interpretable dependence network among environmental, social, and macro-structural indicators. The estimated graph reveals that the empirical organization of sovereign ESG data is not well described by a strict separation of the conventional ESG pillars, but rather by a small number of cross-pillar structural mechanisms, alongside more autonomous  components. The inspection of selected cross-covariance surfaces further enriches this picture by showing not only whether two indicators are linked, but also the sign and temporal persistence of their association. Taken together, these findings suggest that the proposed methodology can provide a useful statistical framework for the study of high-dimensional functional data, where missingness and cross-domain interactions are central empirical features.}

\section*{Code and data availability}

The \texttt{R} script to replicate the simulation study and the data that support the findings of this study are available from
https://github.com/gianluca-sottile/Gaussian-Graphical-Models-for-Partially-Observed-Multivariate-Functional-Data. 

\section*{Acknowledgments}
Luigi Augugliaro and Gianluca Sottile gratefully acknowledge financial support from the University of Palermo (FFR2024 and FFR2025). The authors were also financially supported by the European Union - Next Generation EU - Mission 4 Component 2 - CUP: B53D23009470006 (Marco Borriero), B53D23009480006 (Luigi Augugliaro and Gianluca Sottile) and C53D23002580006 (Veronica Vinciotti). \bt{We thank the anonymous reviewers for their  constructive feedback and insightful suggestions, which greatly improved the quality of the manuscript}
\bibliographystyle{abbrvnat}
\bibliography{fggm}

\appendix

\section{Proof of Theorem~\ref{thm:Bhat}}\label{appendixA}

To obtain the system~(\ref{sistema}), we start rewriting the objective functional, denoted by $f(\bm B_k)$, in a more convenient way. First, we notice that it is equal to the sum of three specific terms:

\begin{equation}\label{eqn:fB}
\begin{split}
f(\bm B_l) =& \frac{1}{2} \mathbb{E} \{(\bm{\xi^m}_l - \langle \bm B_l, \bm X^{\bm O}\rangle)^\top \Theta_l^{m} (\bm{\xi^m}_l - \langle \bm B_l, \bm{X^O}\rangle) \}\\
=& \frac{1}{2} \mathbb{E}[\mathrm{tr}\{ \Theta_l^{m} (\bm{\xi^m}_l - \langle \bm B_l, \bm{X^O}\rangle)(\bm{\xi^m}_l - \langle \bm B_l, \bm{X^O}\rangle)^\top\}]\\
=&  \frac{1}{2} \mathrm{tr}[\Theta_l^{m} \mathbb{E} \{(\bm{\xi^m}_l)(\bm{\xi^m}_l)^\top\}] + \frac{1}{2} \mathrm{tr}[\Theta_l^{m} \E\{\langle \bm B_l, \bm{X^O}\rangle\langle \bm B_l, \bm{X^O}\rangle^\top\}] \\
& - \mathrm{tr}[\Theta_l^{m} \mathbb{E} \{\bm{\xi^m}_l \langle\bm B_l, \bm {X^O}\rangle^\top\}] 
\end{split}
\end{equation}

\noindent The first term in~(\ref{eqn:fB}) can be further simplified by recalling that $\xi^m_{hl} = \langle X_h^{M_h}, \varphi^{M_h}_l\rangle$, from which
\[
\mathbb{E}(\xi^m_{hl} \xi^m_{kl}) = \mathbb{E}(\langle X_h^{M_h}, \varphi^{M_h}_l\rangle \langle X_k^{M_k}, \varphi^{M_k}_l\rangle) = \left\langle\left(\langle \mathbb{E}(X_h^{M_h} X_k^{M_k}),\varphi^{M_k}_l\rangle\right),\varphi^{M_l}_l\right\rangle = \langle\mathcal C_{hk}^{M_hM_k}\varphi^{M_k}_l,\varphi^{M_h}_l\rangle,
\]
where $\mathcal C_{hk}^{M_hM_k}$ denotes the cross-covariance operator with kernel $C_{hk}^{M_hM_k}(s, t) = \mathbb{E}\{X_h^M(s)X_k^M(t)\}$. Using the previous identity we have
\begin{equation}\label{eqn:fB_1}
\frac{1}{2} \mathrm{tr}[\Theta_l^{m} \mathbb{E} \{(\bm{\xi^m}_l)(\bm{\xi^m}_l)^\top\}] = %
\frac{1}{2}\sum_{h = 1}^p\sum_{k = 1}^p\theta_{hkl}^{m} \mathbb{E}(\xi^m_{hl} \xi^m_{kl}) = 
\frac{1}{2}\sum_{h = 1}^p\sum_{k = 1}^p\theta_{hkl}^{m} \langle\mathcal C_{hk}^{M_hM_k}\varphi^{M_k}_l,\varphi^{M_h}_l\rangle,
\end{equation}
which shows that the first term in $f$ is a constant with respect to $\bm B_l$, and therefore can be removed from our initial problem.

To rewrite the second term in~(\ref{eqn:fB}) in a more straightforward way, we notice that the generic term of the matrix $\E\{\langle \bm B_l, \bm{X^O}\rangle\langle \bm B_l, \bm{X^O}\rangle^\top\}$ is equal to:
\begin{eqnarray*}
\E\left\{ \langle \bm \beta_{hl}, \bm{X^O}\rangle\langle \bm \beta_{kl}, \bm {X^O}\rangle\right\} &=& \sum_{i = 1}^p\sum_{j = 1}^p\E[\langle\beta_{ihl},  X^{O_i}_i\rangle\langle\beta_{jkl},  X^{O_j}_j\rangle] \\
&=& \sum_{i = 1}^p \langle\beta_{ihl},\langle\bm{C}^{OO}_i,\bm\beta_{kl}\rangle \rangle \\
&=& \langle\bm\beta_{hl},\langle\bm{C}^{OO},\bm\beta_{kl}\rangle \rangle \\
&=& \langle\bm\beta_{hl},\mC^{OO}\bm\beta_{lk}\rangle.
\end{eqnarray*}
Therefore we have:
\begin{eqnarray}\label{eqn:fB_2}
\frac{1}{2} \mathrm{tr}[\Theta_l^{m} \E\{\langle \bm B_l, \bm{X^O}\rangle\langle \bm B_l, \bm{X^O}\rangle^\top\}] &=& \frac{1}{2} \sum_{h=1}^p\sum_{k=1}^p \theta_{hkl}^{m} \langle\bm\beta_{hl},\mC^{OO}\bm\beta_{kl}\rangle\nonumber\\
&=& \frac{1}{2} \mathrm{tr}[\Theta_l^m \langle \bm B_l, \mC^{OO}\bm B_l\rangle],
\end{eqnarray}
where, with a slight abuse of notation,  we denote by $\mC\bm B_l$ and $\langle \bm B_l, \mC^{OO}\bm B_l\rangle$ the following matrices:
\[
\mC^{OO}\bm B_l = (\mC^{OO}\bm\beta_{1l}\mid\cdots\mid \mC^{OO}\bm\beta_{pl})\quad\text{and}\quad \langle \bm B_l, \mC^{OO}\bm B_l\rangle = %
\begin{pmatrix}
\langle\bm{\beta}_{1l}, \mC^{OO}\bm\beta_{1l} \rangle & \cdots & \langle\bm{\beta}_{1l}, \mC^{OO}\bm\beta_{pl} \rangle \\
\vdots & \ddots & \vdots \\
\langle\bm{\beta}_{pl}, \mC^{OO}\bm\beta_{1l} \rangle & \cdots & \langle\bm{\beta}_{pl}, \mC^{OO}\bm\beta_{pl} \rangle
\end{pmatrix}.
\]
Finally, we focus on the last term in~(\ref{eqn:fB}). Using the identity $\xi^m_{hl} = \langle X_h^{M_h}, \varphi^{M_h}_l\rangle$, the generic term in $\mathbb{E} \{\bm{\xi^m}_l \langle\bm B_l, \bm {X^O}\rangle^\top\}$ can be written as follows
\begin{eqnarray*}
\E\{\xi_{hl}^m \langle\bm \beta_{kl}, \bm {X^O}\rangle\} &=& \E\{\langle X_h^{M_h}, \varphi^{M_h}_l\rangle \langle\bm \beta_{kl}, \bm {X^O}\rangle\} \\
&=& \sum_{i = 1}^p\E\{\langle X_h^{M_h}, \varphi^{M_h}_l\rangle \langle\beta_{ikl}, X^{O_i}_i\rangle\}\\
&=& \sum_{i = 1}^p \langle \beta_{ikl}, \langle C^{O_iM_h}_{ih},  \varphi_l^{M_h}\rangle\rangle \\
&=& \langle \bm\beta_{kl}, \bm R_{hl}\rangle,
\end{eqnarray*}
where $\bm R_{hl}$ is the $h$th column of the matrix
\[
\bm R_l = (\bm R_{1l}\mid\cdots\bm R_{pl}) = %
\begin{pmatrix}
\mC_{11}^{O_1M_1} \varphi^{M_1}_l & \dots & \mC_{1p}^{O_1M_p} \varphi^{M_p}_l \\
\vdots & & \vdots \\
\mC_{p1}^{O_pM_1} \varphi^{M_1}_l & \dots & \mC_{pp}^{O_pM_p} \varphi^{M_p}_l
\end{pmatrix}.
\]
Using the previous identities we have that the third term in~(\ref{eqn:fB}) can be written as follows:
\begin{equation}\label{eqn:fB_3}
\mathrm{tr}[\Theta_l^{m} \mathbb{E} \{\bm{\xi^m}_k \langle\bm B_l, \bm {X^O}\rangle^\top\}] = \sum_{h=1}^p\sum_{k=1}^p \theta_{hkl}^{m} \langle \bm\beta_{kl}, \bm R_{hl}\rangle = \mathrm{tr}[\Theta_l^{m}\langle \bm B_l, \bm R_l\rangle].
\end{equation}
Given the identities~(\ref{eqn:fB_1}), (\ref{eqn:fB_2}) and (\ref{eqn:fB_3}), we can write the objective functional as follows
\[
f(\bm B_l) = \frac{1}{2} \mathrm{tr}[\Theta_l^M \langle \bm B_l, \mC^{OO}\bm B_l\rangle] - \mathrm{tr}[\Theta_l^{m}\langle \bm B_l, \bm R_l\rangle] + Cost.
\]
From this, taking the Fréchet derivative of $f(\bm B_l)$ with respect to $\bm B_l$ and setting it equal to zero we have
\[
\frac{\partial f(\bm B_l)}{\partial \bm B_l} =  \Theta_l^{m} \mathcal{C}^{O O}\bm B_l -  \Theta_l^{m}\bm{R}_l=\bm 0.
\]
By multiplying both sides on the left by $\left(\Theta_l^{m}\right)^{-1}$ we obtain the following system
\[
\mathcal{C}^{O O}\bm B_l - \bm{R}_l=\bm 0,
\]
in the unknown $\bm B_l$.

\section{Computational cost}\label{supp:comp_cost}

\bt{Computational times were recorded on a 2024 iMac equipped with an Apple M4 chip (10 cores: 4 performance and 6 efficiency cores) and 16 GB of RAM, running macOS 26.2. The simulations were carried out using the \texttt{R} version 4.4.3 on the \texttt{aarch64-apple-darwin20} platform. Matrix computations relied on the default BLAS and LAPACK libraries provided by the system and by the \texttt{R} installation. The implementation of the proposed method combines \texttt{R} and \texttt{Rcpp/C++} code. The reported computational times correspond to wall-clock elapsed time on this specific hardware/software configuration and are intended primarily as relative comparisons across simulation settings.}

\end{document}